\documentclass[final,5p,times,twocolumn,authoryear]{elsarticle}

\usepackage{graphicx}	
\usepackage{amsmath}	
\usepackage{amssymb}	
\usepackage{bm}         
\usepackage{ulem}       
\usepackage{verbatim}   
\usepackage{newtxtext,newtxmath}
\usepackage{lipsum}
\usepackage{hyperref}

\newcommand{\python}{\texttt{Python}}
\newcommand{\reb}{\texttt{REBOUND}}
\newcommand{\mercurius}{\texttt{MERCURIUS}}
\newcommand{\Nbody}{\textit{N}-body}

\journal{New Astronomy}

\begin{document}

\begin{frontmatter}

\title{Composition Tracking for Collisions Between Differentiated Bodies in REBOUND}
    
\author[unlv]{Noah Ferich\corref{cor1}}
\cortext[cor1]{Corresponding Author}
\ead{fericn1@unlv.nevada.edu}
\author[NU]{Anna C. Childs}
\author[unlv]{Jason H. Steffen}
\ead{jason.steffen@unlv.edu}
\affiliation[unlv]{organization={Department of Physics \& Astronomy, University of Nevada, Las Vegas},
            addressline={4505 S. Maryland Pkwy}, 
            city={Las Vegas},
            postcode={89154}, 
            state={Nevada},
            country={USA}}

\affiliation[NU]{organization={Center for Interdisciplinary Exploration and Research in Astrophysics (CIERA) and Department of Physics and Astronomy, Northwestern University},
            addressline={1800 Sherman Ave}, 
            city={Evanston},
            postcode={60201}, 
            state={Illinois},
            country={USA}}

\begin{abstract}
Previous research suggests that impacts between planetary embryos and planetesimals during the late stages of planet formation can often determine the percentages of core and mantle material that compose the newly formed planets in a system. Previous studies have attempted to include the composition-changing effects of these collisions in N-body simulations of planet formation, often as post-processing codes. In this paper, we present the Differentiated Body Composition Tracker, a new post-processing tool that uses collisional data collected from the N-body integrator REBOUND to determine the amount of core and mantle material that is transferred between colliding objects and the resulting fragments during an impact. We demonstrate how this code works using the data from 50 REBOUND simulations of planet formation and explore how the parameters in the code affect the core mass fractions of the remaining objects from these simulations. We then investigate how non-uniform distributions of core material across an initial disc affect the final core mass fractions of planets. Under ideal conditions, we find that a combination of giant impacts and planetary embryos enriched in core material could create some of the iron-rich planets that have been discovered. 
\end{abstract}


\end{frontmatter}

\section{Introduction}
\label{sec:intro}

Low-mass, rocky exoplanets are abundant in the galaxy and the main candidates for habitable worlds \citep{Fressin2013, Bryson2021}. Similar to the rocky planets of the Solar System, many of these exoplanets are believed to contain predominantly silicate mantles and iron-rich cores.  The ratio of core to mantle material in planets may be linked to the the compositions of their host stars \citep{Bonsor2021}, yet current measurements suggest that the core mass fractions (CMFs) of many exoplanets can vary significantly from what's expected from stellar metallicities \citep{Plotkynov2020}.  Even though the sample size of observed rocky exoplanets is relatively small, the range between their measured CMFs is broad and can reach values below 0.10 and above 0.70 (See Figure 1 of \citet{Scora2022}).  Examples of exoplanets with high CMFs include Kepler-105c with a CMF of approximately 0.72 \citep{Jontof-Hutter2016, Hadden2017}, Kepler-107c with a CMF of approximately 0.67-0.70 \citep{Bonomo2019, Schulze2021}, and K2-38b with a CMF of approximately 0.68 \citep{Toledo-Padron2020}. Furthermore, not only do the CMFs of exoplanets vary between systems, but can also vary within the same system, with Mercury (CMF$\approx$0.70 \citep{Huack2013}) and Earth (CMF$\approx$0.33 \citep{Stacey2005}) being two local examples. These compositional discrepancies open questions about the processes that formed these planets.

Previous studies have explored how planets obtain CMFs that are substantially higher than expected from their host star's metallicity \citep{Plotkynov2020}. Recent work done by \citet{Aguichine2020} and \citet{Johansen2022} details different mechanisms that can lead to the formation of iron-rich plantesimals close to stars, which accrete into planets with high CMFs like Mercury. While these newer theories are promising, future work needs to constrain the properties of planets produced by these mechanisms. In addition, \citet{Santerne2018} suggested that part of the mantle could be lost through the photo-evaporation of silicate materials. However, this mechanism only significantly affects planets with extremely short periods and is unlikely to remove a considerable amount of mantle material for planets with masses larger than Earth \citep{Santerne2018, Ito2021}.

The majority of works suggest that this compositional diversity results from giant impacts between planetary embryos and planetesimals in the late stages of planet formation \citep{Asphaug2006}. Smoothed Particle Hydrodynamics (SPH) simulations show how single impacts between differentiated planetary bodies can strip away large portions of mantle material, leaving behind planets that are enriched in iron \citep{Carter2018, Reinhardt2022}. While one impact may not be enough to explain some of the higher CMFs seen in exoplanets and Mercury \citep{Franco2022}, multiple impacts during a planet's formation could possibly lead to the CMFs that are observed.

Multiple studies have used \Nbody\ simulations to study the formation and evolution of planetary systems, but have only recently started to include imperfect collisions and fragmentation in these simulations \citep{Kokubo2010, Chambers2013, Bonsor2015, Childs2019, Scora2020, Kaufman2023}.  While many of these studies focus on how these imperfect collisions can change the compositions of newly formed planets \citep{Chambers2013, Bonsor2015, Carter2015, Childs2019, Scora2020}, questions surrounding the formation process and how exoplanets can have such a wide range of elemental abundances remain unanswered. These issues will become increasingly relevant as future studies and missions collect more data on the interior compositions of exoplanets. Therefore, the community will benefit from having a wide range of tools at their disposal to study these problems.

Recently, \citet{Childs2022} used the semi-analytic collision model from \citet{Leinhardt2011a} to create a fragmentation module for the \Nbody\ integrator \reb\ \citep{Rein2012}. This module allows users to include imperfect collisions between objects in their simulations. In addition, they created a post-processing composition tracking code that uses data collected from this module to track how material is exchanged between objects and fragments during collisions. While this code can provide a good approximation for how different materials can be transferred after impacts, it assumes that all objects are homogeneous and completely undifferentiated, an assumption that prevents the code from being able to model the uneven transfer of core and mantle material between differentiated bodies.

Previous models have attempted to track this exchange of core and mantle material between colliding objects and their remnants.  To estimate this process, \citet{Marcus2010-1} created two basic prescriptions that assume extreme net losses or gains in core material for objects after impact. Approximations such as these can be improved with the use of SPH simulations; however, performing these calculations for every collision in an \Nbody\ simulation is computationally taxing and currently unfeasible for most simulations of any significant length. To alleviate this issue, numerous studies have used suites of SPH simulations to create semi-analytic \citep{Marcus2009, Carter2018, Reinhardt2022} or machine learning models \citep{Timpe2020, Winter2023} that can determine the final compositions and other properties of remnants from collisions.

In this paper, we introduce a new post-processing code that tracks the composition of differentiated bodies from \reb\ simulations. It contains a prescription for the transfer of mantle and core material between objects during a collision depending on the collision type, the amount of mass exchanged during the impact, the core and outer radii of the colliding objects, and the impact parameter of the collision (more details provided in \S~\ref{sec:DBCT}). Adding an additional layer to the composition tracking code in \citet{Childs2022} will allow objects to preferentially lose or gain one type of material over the other during collisions, which can drastically alter their CMFs.

The fragmentation module, like the composition tracking code from \citet{Childs2022}, assumes that objects contain a single layer; therefore, our post-processing code must make certain assumptions (see \S~\ref{sec:implementation}) about the objects in these \reb\ simulations when performing its calculations. Our goal is to provide a framework to estimate how the compositions of these objects change if we assume that the collisions produced by the fragmentation module were between differentiated bodies. In doing so, we may better evaluate the importance of incorporating differentiation in planet formation models. Many of the assumptions in our framework can be modified and the framework itself can be updated in the future with results from SPH simulations to give it more detailed modeling capabilities. 

We use our Differentiated Body Composition Tracker (DBCT) in conjunction with the \reb\ fragmentation module to investigate how the final CMFs of planets vary depending on the initial conditions and collisional history of the late-stage protoplanetary disc from which they formed. In Section \ref{sec:rebound_frag_and_comp_track}, we briefly discuss the \reb\ fragmentation module and the composition tracking code created by \citet{Childs2022}. In Section \ref{sec:DBCT}, we discuss the model behind the DBCT and how we implement it as a post-processing code that uses data from the fragmentation module. In Section \ref{sec:nbody_simulations}, we discuss the simulation data that we use with our DBCT for all the results produced in this paper. In Section \ref{sec:uniform_disc}, we use our DBCT on an initial disc with uniform CMFs and discuss various findings surrounding the results. In Section \ref{sec:varying_disc_composition}, we investigate how the final compositions of planets change depending on the arrangement of core material in the initial disc. We then compare these results to current exoplanet data and see if gradients of core material in the disc can explain the high planetary CMFs that have been observed. In Section \ref{sec:Discussion} and \ref{sec:conclusion}, we summarize the findings from all previous sections and describe topics to investigate in future studies using our code.

\section{REBOUND Fragmentation and Composition Tracking} 
\label{sec:rebound_frag_and_comp_track}
\subsection{Fragmentation Module}

The fragmentation module uses the works of \citet{Leinhardt2012}, \citet{Asphaug2010}, and \citet{Genda2012} to determine the outcome of collisions during simulations and has been implemented into \reb\ using methods created by \citet{Chambers2013}. When two objects collide, the larger of the two is considered the target and the smaller object is considered the projectile. The module uses the angle of impact, impact velocity, mass and radius of the target, and mass and radius of the projectile to resolve the collision and determine how mass is transferred between the two objects and into newly created fragments. Users have the option to set the minimum mass that a fragment can receive, which often determines the number of fragments that form in a collision or if fragments are produced at all. These fragments are given the same density as the target and are placed in an equidistant circle centered around the largest remaining object from the collision. Depending on the collision type, the projectile can remain intact and become the second largest remnant from the collision, be destroyed and have its mass deposited into these fragments, or be absorbed entirely into the target.

In total, the fragmentation model has eight different collision outcomes that can occur: elastic bounce, perfect merger, effective merger, graze-and-merge, partial accretion, hit-and-run, partial erosion, and super-catastrophic. In an elastic bounce, both the target and projectile remain unchanged and no mass is transferred between them. In perfect mergers, effective mergers, and graze-and-merges, all of the mass in the projectile aggregates onto the target and the projectile is removed from the simulation. The remaining object is the largest remnant and is the only particle that comes out of these types of collisions. In partial accretions and hit-and-runs, a certain fraction of the projectile's mass is transferred to the target. The target again becomes the largest remnant and will always have a mass that is greater than or equal to the initial mass of the target. If the collision is a partial accretion, the projectile is removed from the simulation and its remaining mass is placed into fragments; if the collision is a hit-and-run, there is a possibility (depending on the collision parameters) that the projectile will remain intact and become the second largest remnant. In partial erosion and super-catastrophic collisions, the projectile strips a certain amount of mass off of the target. The target becomes the largest remnant again and will always have a mass that is less than the initial mass of the target going into the collision. In both cases, the projectile is destroyed; its mass and the mass ejected from the target are placed into fragments. A super-catastrophic collision is defined as a partial erosion collision where the largest remnant is less than or equal to one-tenth of the initial mass of the target. See \citet{Childs2022} for greater detail about each collision regime and the mathematics behind the fragmentation module.

\subsection{Composition Tracking}

As mentioned in \S~\ref{sec:intro}, \citet{Childs2022} also created a \python\ post-processing code that takes collisional data collected with the fragmentation module and tracks how materials are transferred between objects after impacts. Users first create a file containing the unique identifier for each initial object in the simulation (known as a hash in \reb), the mass of each body, and the starting fractional abundances of each material in the objects. The code uses this compositional input file and the collision report from a \reb\ simulation and returns an output file containing the final compositions for the remaining objects in the same format as the compositional input file. 

While the code allows users to track an arbitrary number of constituents, it assumes that all objects are homogeneous and undifferentiated. This assumption is limited in its application because small planetesimals can undergo differentiation in a few million years or less \citep{Kruijer2014, Neumann2014, Lichtenberg2019, Carry2021} and planetary embryos can differentiate between collisions \citep{Dahl2010}. To get a more accurate sense of how material is transferred between objects during collisions, the separation of an object's compositional species into a distinct core and mantle must be taken into account.

\section{Differentiated Body Composition Tracker}
\label{sec:DBCT}
\subsection{Model}
\label{sec:model}

We present a model for the transfer of core and mantle materials between two colliding objects. This model determines the fraction of core material in the ejecta of a collision based on the impact parameter (see equation \ref{eq:b}), the core radius and outer radius of the target and projectile, and the minimum and maximum fraction of core material that can be in the ejecta. These last two parameters can be set by users when using the model.

\begin{figure}
	\includegraphics[width=\columnwidth]{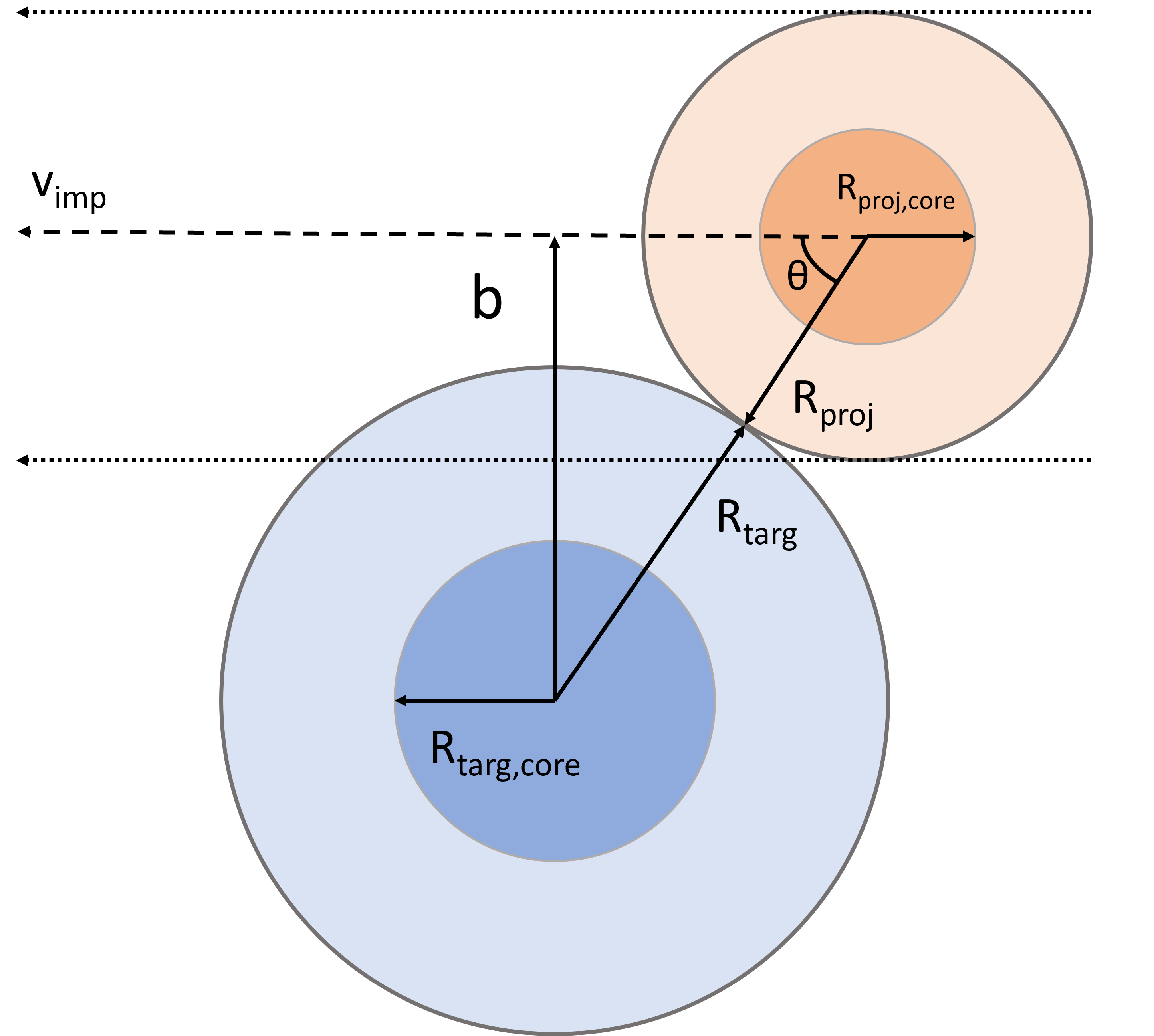}
    \caption{Two-dimensional diagram of a partial erosion collision between two differentiated objects. The dotted lines show how the cross-section of the projectile will move through the target if it continues on its current path.}
    \label{fig:impact_diagram}
\end{figure}

Fig. \ref{fig:impact_diagram} shows a diagram of a collision between two, spherical objects at the moment of impact. Both the target and projectile are differentiated and are composed of an inner, spherical layer of dense core material and an outer spherical shell of less dense mantle material. $R\textsubscript{targ, core}$ is the radius of the target's core, $R\textsubscript{targ}$ is the target's outer radius, $R\textsubscript{proj, core}$ is the radius of the projectile's core, and $R\textsubscript{proj}$ is the projectile's outer radius. $v\textsubscript{imp}$ is the velocity of the impact and is given by the following equation:
\begin{equation}
v\textsubscript{imp} = \sqrt{v\textsubscript{rel}^2-2GM\textsubscript{tot}\left(\frac{1}{x\textsubscript{rel}}-\frac{1}{R\textsubscript{tot}}\right)};
\label{eq:v_imp}    
\end{equation}
where $v\textsubscript{rel}$ is the relative velocity between the two objects, $G$ is the gravitational constant, $M\textsubscript{tot}$ is the total mass of the two bodies, $x\textsubscript{rel}$ is the distance between the centers of the objects, and $R\textsubscript{tot}$ is the sum of the target's and projectile's radius. Fig. \ref{fig:impact_diagram}, shows the target as stationary with $v\textsubscript{imp}$ simply being the velocity of the projectile upon impact. 

Fig. \ref{fig:impact_diagram} also shows $\theta$, the angle that subtends the line connecting the centers of the two objects and their relative velocity vector at the moment of impact, and $b$, the impact parameter, which is used extensively in our DBCT and can be calculated with
\begin{equation}
b = (R\textsubscript{targ}+R\textsubscript{proj})\sin{\theta}.
\label{eq:b}    
\end{equation}
\begin{figure}
\includegraphics[width=\columnwidth]{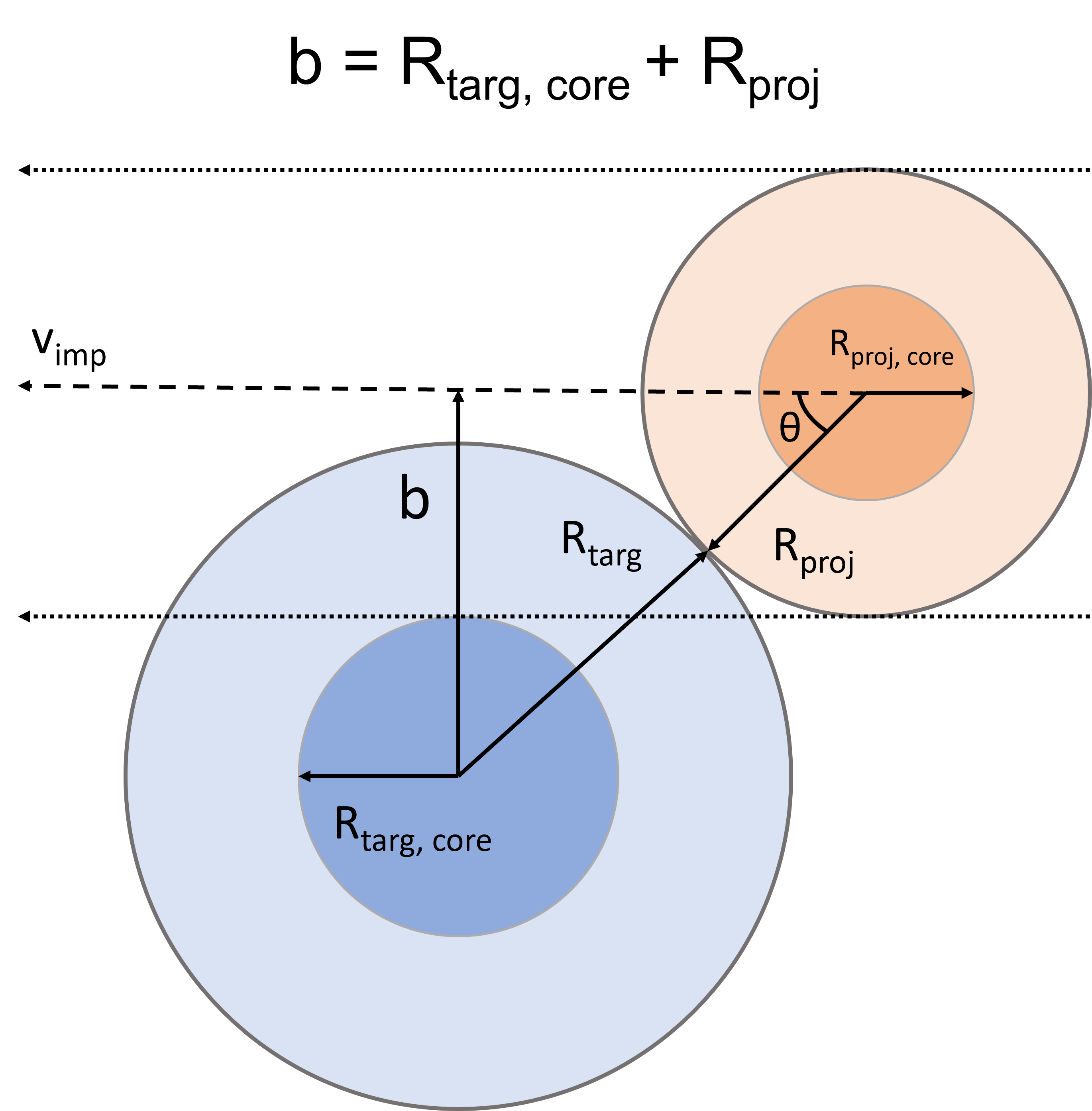}
    \caption{Diagram of a partial erosion collision. When $b=R\textsubscript{proj}+R\textsubscript{targ,core}$, the cross-section of the projectile will pass by the edge of the target's core. According to our DBCT model, as the impact parameter decreases below this value, the target starts losing a larger fraction of core material during the collision. If $b$ is at or above this value, the target loses the minimum amount of core material during a collision.}
    \label{fig:max_impact_diagram}
\end{figure}
The dotted lines in Fig. \ref{fig:impact_diagram} show how the cross-section of the projectile will pass through the target if it continues along its current trajectory after the moment of impact. In Fig. \ref{fig:impact_diagram}, the cross-section of the projectile does not pass through the target's core, and in our model, we consider this to be a partial erosion collision where the fraction of core material in the collisional ejecta is at a minimum. When the impact parameter decreases below $b=R\textsubscript{proj}+R\textsubscript{targ,core}$, the cross-section of the projectile will pass through the target's core and the fraction of core material in the ejecta will increase above this minimum. Fig. \ref{fig:max_impact_diagram} shows where this threshold occurs.
\begin{figure}
\includegraphics[width=\columnwidth]{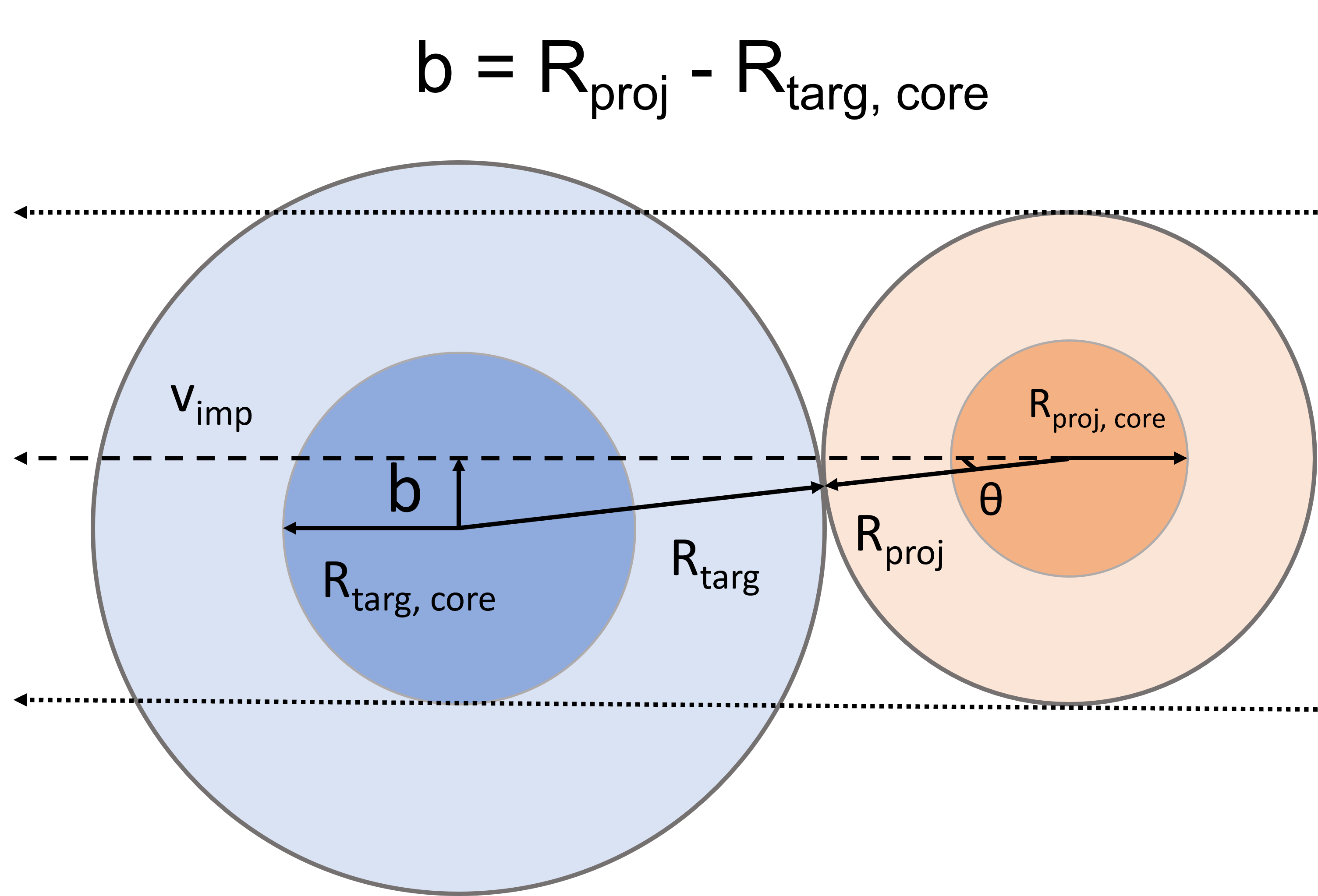}
    \caption{Diagram of a partial erosion collision. When $b=R\textsubscript{proj}-R\textsubscript{targ,core}$, the cross-section of the projectile will pass through the entire core of the target. According to our DBCT model, if the value of $b$ reaches or drops below this value, the target loses the the maximum amount of core material from the collision. If $b$ is above this value, then the amount of core material eroded from the target will start to decrease.}
    \label{fig:min_impact_diagram}
\end{figure}

As $b$ decreases beyond the boundary shown in Fig. \ref{fig:max_impact_diagram}, the fraction of core material in the ejecta will linearly increase until it reaches its maximum value when the target's core lies fully within the projectile's cross-section. This occurs when the impact parameter reaches or surpasses a value of $b=R\textsubscript{proj}-R\textsubscript{targ,core}$ and is shown in Fig. \ref{fig:min_impact_diagram}.

Under these parameters and regimes, the model for our DBCT for partial erosion collisions and super-catastrophic collisions can be described mathematically as
\begin{equation}
\label{eq:erosion_model}
    F\textsubscript{core} =  \begin{cases}
        f\textsubscript{min}, &  b \geq R\textsubscript{proj}+R\textsubscript{targ,core}\\
         f\textsubscript{max}, &  b \leq R\textsubscript{proj}-R\textsubscript{targ,core}\\
         f\textsubscript{min}+s\textsubscript{e}(b-(R\textsubscript{proj}+R\textsubscript{targ,core})), &  \text{otherwise}.
    \end{cases}.
\end{equation}
Our DBCT code uses the impact parameter of the collision, inner radii of the colliding objects, and outer radii of the colliding objects to calculate $F\textsubscript{core}$, the CMF of the collision ejecta. $f\textsubscript{min}$ and $f\textsubscript{max}$ are user-entered variables that determine the minimum and the maximum fraction of core material that composes the ejecta. $s\textsubscript{e}$ is the slope of the erosive collision model and is used to determine the CMF of the ejecta when the cross-section of the projectile partially passes through the target's core. It is described by the following equation:
\begin{equation}
 s\textsubscript{e} = -\frac{f\textsubscript{max}-f\textsubscript{min}}{2R\textsubscript{targ,core}}.
\label{eq:slope_pe}   
\end{equation}
The DBCT model for collisions where the projectile looses mass (accretive) is similar to the model where the target looses mass (erosive). However, the model for accretive collisions depends on where the cross-section of the target is in comparison to the projectile's core and mantle at the moment the projectile impacts the target. This model can be described by the following equation:
\begin{equation}
\label{eq:accretion_model}
    F\textsubscript{core} = \begin{cases}
        f\textsubscript{min}, &  b \geq R\textsubscript{targ}+R\textsubscript{proj,core}\\
         f\textsubscript{max}, &  b \leq R\textsubscript{targ}-R\textsubscript{proj,core}\\
         f\textsubscript{min}+s\textsubscript{a}(b-(R\textsubscript{targ}+R\textsubscript{proj,core})), &  \text{otherwise}
    \end{cases};
\end{equation}
where $s\textsubscript{a}$ is the slope for the linear part of this model:
\begin{equation}
 s\textsubscript{a} = -\frac{f\textsubscript{max}-f\textsubscript{min}}{2R\textsubscript{proj,core}}.
\label{eq:slope_pa}   
\end{equation}
This model is flexible and has parameters that can be updated with results from SPH simulations. More specifically, SPH calculations can be used to determine what $f\textsubscript{min}$ and $f\textsubscript{max}$ should be depending on the parameters of the collision and the physical properties of the colliding bodies. The SPH results can also be used to perform a more detailed calculation of how much core material should be in the impact ejecta.

\subsection{Implementation}
\label{sec:implementation}

To use the DBCT, users need a collision report from the fragmentation module after the simulation's completion and must create a compositional input file containing the hashes, masses, and CMFs of all the initial objects in the corresponding \reb\ simulation. The code outputs a file with the hashes, masses, and CMFs of the remaining objects. We focus on tracking the fraction of core and mantle material contained in an object, and therefore, the code is limited to modelling objects that are composed of two distinct layers of material. We make the simplifying assumption that objects are always fully differentiated -- even directly after collisions -- and that objects are perfectly spherical with a spherical core and spherical shell of mantle material around it.

We have modified the fragmentation module of \citep{Childs2022} to output all of the parameters needed for our DBCT to work. The collision reports created by the standard version of the fragmentation module give the time of the collision, the collision type, the \reb\ hash of the target, the \reb\ hash of the projectile, the mass of the largest remnant, and the hashes and masses of fragments (if they were created). Our version outputs all of these parameters and additionally gives the sine of the impact angle, which we will use to calculate the impact parameter used in the DBCT for each collision, and the outer radii of the target and projectile. This updated fragmentation module will be made available with our DBCT\footnote{See Data Availability}. 

The starting files do not contain the core radius and outer radius of each object. Because the fragmentation module assumes that objects are uniform and have a constant bulk density, we must obtain an estimate for the core radius and outer radius for both the target and projectile involved in each collision.  To do this, we first calculate what the core radius, $r\textsubscript{core}$, will be for an object with mass $M$ and a CMF of $F\textsubscript{core}$:
\begin{equation}
r\textsubscript{core} = \left(\frac{3MF\textsubscript{core}}{4\pi\rho\textsubscript{core}}\right)^\frac{1}{3}.
\label{eq:R}    
\end{equation}
where $\rho\textsubscript{core}$ is the uncompressed density of the core material and is a parameter that can be changed by users. We assume the core of each object is made entirely of iron and therefore give $\rho\textsubscript{core}$ a value of 7874.0 kg m$^{-3}$, the density of iron. We then calculate the outer radius, $r$, for this object with the following equation: 
\begin{equation}
r = \left(R\textsubscript{core}^3+\frac{3M}{4\pi\rho}\right)^\frac{1}{3},
\label{eq:R_core}    
\end{equation}
where $\rho$\textsubscript{mantle} is the density of the mantle material and is also a parameter that can be changed by users. We give it a value of 3000.0 kg m$^{-3}$, which is the density value used to calculate $R\textsubscript{sim}$, the radius of objects in the simulations we use (see \S~\ref{sec:nbody_simulations}). In summary, bodies in the DBCT comprise a spherical, purely iron core and a spherical, purely mantle shell. While more detailed models of the interior of a rocky object could be used, this approximation was chosen to the work with the model described in \S~\ref{sec:model}.
 
With this implementation, $r$ will only match $R\textsubscript{sim}$ when an object has a CMF of 0.0. If an object has a CMF that is greater than 0.0, than $r$ will always be smaller than $R\textsubscript{sim}$ due to the higher density of core material\footnote{Here we ignore the effects of compression.}. The radius discrepancy between the differentiated objects in the post-processing code and the all-mantle bodies in these simulations is one downside to performing composition tracking calculations after the completion of these simulations. Varying the CMFs of objects during the simulation will change their outer radii, which will then change the parameters and outcomes of certain collisions and the evolution of the planetary system itself. Because the objects in these simulations are composed of only mantle material, the radii of each body is maximized, which will lead to a higher collisional cross-section and higher chance of the object undergoing an impact. Therefore, we assume that these simulations have a larger number of collisions than average.

In addition to the core and outer radius of each object, we must obtain an estimate for the impact parameter of the collision using the new radii:
\begin{equation}
b\textsubscript{calc} = (r\textsubscript{targ}+r\textsubscript{proj})\sin{\theta},
\label{eq:b_calc}    
\end{equation}
where $r\textsubscript{targ}$ is the outer radius of the target as calculated by the DBCT, $r\textsubscript{proj}$ is the outer radius of the projectile as calculated by the DBCT, and $b\textsubscript{calc}$ is the impact parameter that we will use with the DBCT for each collision.

The DBCT iterates through each collision in the collision report, determining the post-impact CMFs for objects based on the parameters of the impact, properties of the colliding objects, and the collision type of the impact.  If the collision is an elastic bounce, the code moves on to the next collision. If the collision is a merger, then the largest remnant's CMF is calculated as:
\begin{equation}
F\textsubscript{LR} = \frac{(M\textsubscript{targ}F\textsubscript{targ})+(M\textsubscript{proj}F\textsubscript{proj})}{M\textsubscript{LR}},
\label{eq:merger_CMF}    
\end{equation}
where $M\textsubscript{targ}$ is the mass of the target, $F\textsubscript{targ}$ is the CMF of the target, $M\textsubscript{proj}$ is the mass of the projectile, $F\textsubscript{proj}$ is the CMF of the projectile, $M\textsubscript{LR}$ is the mass of the largest remnant, and $F\textsubscript{LR}$ is the largest remnant's CMF. The fragmentation module calculates these post-collision masses, so the differentiated bodies in this model are subject to mass constraints from the simulation data. If the collision is a hit-and-run or partial accretion, then equation \ref{eq:accretion_model} is implemented, and the largest remnant will ideally obtain the following CMF:
\begin{equation}
F\textsubscript{LR} = \frac{(M\textsubscript{targ}F\textsubscript{targ})+(M\textsubscript{ej}F\textsubscript{ej})}{M\textsubscript{LR}},
\label{eq:accretion_CMF_LR}    
\end{equation}
where $M\textsubscript{ej}$ is the mass of the material ejected from the projectile and $F\textsubscript{ej}$ is the ideal CMF of the ejecta that was calculated with equation \ref{eq:accretion_model}. There may be instances where the projectile does not have enough core material to put into the ejecta to meet this ideal CMF.  In this case, the ejecta obtains all of the projectile's core material and extra mantle material will be added to the ejecta so that it achieves the necessary mass. There may also be times when there is not enough mantle material in the projectile to meet this ideal CMF. In this case, the ejecta obtains all of the projectile's mantle material and extra core material will be added to the ejecta so that it achieves the necessary mass. Because of these two possibilities, the ejecta in a collision may have a different CMF than the one calculated using equation \ref{eq:accretion_model}, which will change what the largest remnant's CMF will be according to equation \ref{eq:accretion_CMF_LR}.  

The fragments of the collision and the second largest remnant (if one was formed) will ideally receive a CMF of
\begin{equation}
F\textsubscript{frag} = \frac{(M\textsubscript{proj}F\textsubscript{proj})-(M\textsubscript{ej}F\textsubscript{ej})}{M\textsubscript{tot}-M\textsubscript{LR}},
\label{eq:accretion_CMF_frag}    
\end{equation}
where F\textsubscript{frag} is the ideal CMF of the fragments (and possibly the second largest remnant) and $M\textsubscript{tot}$ is the sum of the target's mass and projectile's mass. 

If the collision is a partial erosion or super-catastrophic, then equation \ref{eq:erosion_model} is implemented, and the largest remnant will ideally obtain the following CMF:
\begin{equation}
F\textsubscript{LR} = \frac{(M\textsubscript{targ}F\textsubscript{targ})-(M\textsubscript{ej}F\textsubscript{ej})}{M\textsubscript{LR}},
\label{eq:erosion_CMF_LR}    
\end{equation}
where $M\textsubscript{ej}$ and $F\textsubscript{ej}$ now come from the target. Once again, the largest remnant's CMF may not match the one described in equation \ref{eq:erosion_CMF_LR} and will depend on the amount of core and mantle material that can be eroded from the target. The fragments of the collision will ideally receive a CMF of:
\begin{equation}
F\textsubscript{frag} = \frac{(M\textsubscript{proj}F\textsubscript{proj})+(M\textsubscript{ej}F\textsubscript{ej})}{M\textsubscript{tot}-M\textsubscript{LR}}.
\label{eq:erosion_CMF_frag}    
\end{equation}
The code performs the aforementioned operations for every impact in the collision report, keeping track of the changing CMFs and masses of objects and the creation of fragments. It also keeps track of objects that were destroyed during collisions and removes their entries from the list once the code is finished going through the collision report. After these objects are removed, the hashes, masses, and CMFs of all the remaining objects are then added to an output file. 

The code also has the ability to remove objects that were ejected from the system and objects that collided with bodies that are not considered part of the disc (such as a star or giant planet outside of the disc). A user may have to modify this part of the DBCT depending on how they determined and collected data on ejections from their \reb\ simulations. 

\section{Tests and Results}
\label{sec:tests_and_results}

In this section, we will apply our DBCT to the results from 50 \reb\ simulations that were performed by \citet{Childs2022} for their investigation into how expansion factors, which artificially inflate the particle radii in \Nbody\ simulations, can affect the final architecture of planetary systems. We will first describe the initial conditions that were used for these \Nbody\ simulations before showing a sample of results from our code assuming a CMF of 0.3 for objects in the simulations' initial discs. Using the results from this uniform disc, we will discuss the final CMFs for the remaining objects in these simulations, explore how the use of an expansion factor can affect the final CMFs of simulated planets, and discern how the variables $f\textsubscript{min}$ and $f\textsubscript{max}$ can affect our model's results. We will then apply our DBCT to the same simulations with the assumption that the CMF of an object in the initial disc depends on its original semi-major axis. Finally, we will compare the results obtained with the DBCT to results obtained with the composition tracking code described in \citet{Childs2022} to explore how the inclusion of differentiation changes the final CMFs obtained from these simulations.

\subsection{N-body Simulations}
\label{sec:nbody_simulations}

Similar to \citet{Chambers2001}, \citet{Childs2022} used a disc of small plantesimals and large planetary embryos for their simulations. A distribution such as this one marks the era of planet formation that is dominated by large impacts and is the period of time when the fragmentation model is most valid \citep{Kokubo2000}. This disc comprises 26 mars-sized embryos (r=0.56 $R_{\oplus}$, mass=0.093 $M_{\oplus}$) and 260 moon-sized planetesimals (r=0.26 $R_{\oplus}$, mass=0.0093 $M_{\oplus}$). All objects have a uniform density of 3000 kg m$^{-3}$. The disc orbits a Sun-like star and has a surface density distribution, $\Sigma$, that follows $\Sigma \sim r^{-3/2}$, where $r$ is the distance from the star. Objects in the disc are distributed between 0.35 au and 4.0 au from the host star. The simulation also includes Jupiter and Saturn at their current positions. 

Eccentricities and inclinations for each object in the disc are chosen randomly from a uniform distribution of $e < 0.01$ and $i < 1^\circ$. In addition, the argument of periastron ($\omega$), the mean anomaly ($M$), and the longitude of ascending node ($\Omega$) are chosen randomly from a uniform distribution that goes from $0^\circ$ to $360^\circ$. The \reb\ random generator seed is changed between each simulation to give new initial conditions to all of the objects in the disc. 

\citet{Childs2022} used five different expansion factors (3, 5, 7, 10, and 15) to artificially increase the radii of each particle in the initial disc, and each expansion factor was used for ten different simulations. Artificially increasing the radii of the bodies in \Nbody\ studies decreases the collision timescale and thus decreases the computational time for planetary systems to complete their evolution. Because of this, the simulation time of each simulation depended on the expansion factor, $f$. Simulations where $f=3$ lasted 100 Myr; simulations where $f=5$ and $f=7$ lasted for 10 Myr; and simulations where $f=10$ and $f=15$ lasted for 5 Myr.

In \reb, \citet{Childs2022} used the integrator \mercurius\ for their simulations \citep{Rein2019}. They gave this integrator an initial time-step of six days, which is about a tenth of the period of the innermost orbit in the disc. For the fragmentation module, the minimum fragment mass was set to 0.0047$M_{\oplus}$, around half of the mass of the planetesimals in the disc. This variable will play a large role in determining if collisions are disruptive or mergers. We will discuss the significance of this value and dealing with debris from collisions in the next section and in \S~\ref{sec:Discussion}.

\subsection{Uniform disc}
\label{sec:uniform_disc}

We enter the collisional data from all 50 of the previously described \Nbody\ simulations into our DBCT. For this example, we assume that all objects start out with a CMF of 0.3. In \S~\ref{sec:varying_disc_composition}, we will change this assumption and give objects in the disc different initial CMFs depending on their location within the disc. We also assume that $f\textsubscript{min}$ and $f\textsubscript{max}$ are 0.0 and 1.0 respectively, the largest possible range between these two variables. In \S~\ref{sec:min_and_max_cmf}, we will discuss how these variables affect the final results from simulations.

\subsubsection{Final CMFs}
\label{sec:uniform_disc_final_cmfs}

\begin{figure}
	\includegraphics[width=\columnwidth]{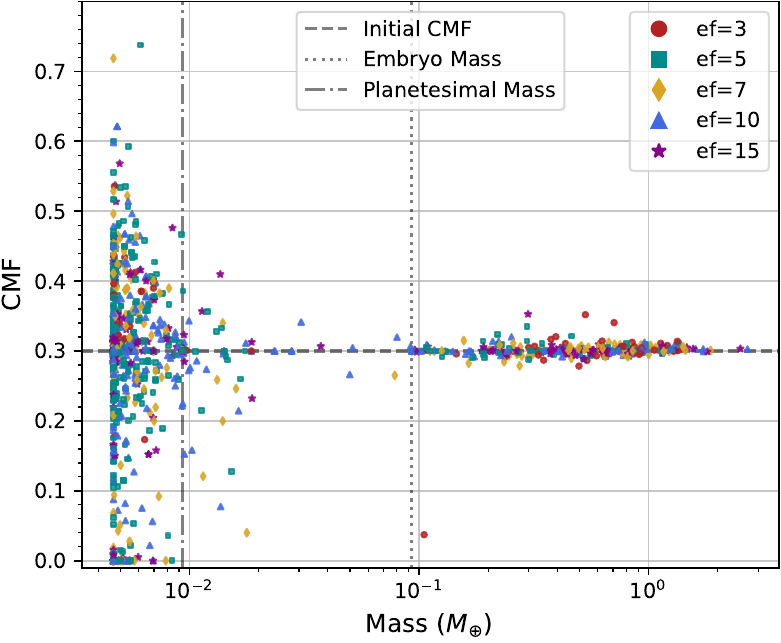}
    \caption{Final Masses and CMFs for the remaining objects in all 50 \reb\ simulations. The horizontal, dashed line shows the starting CMF for all initial objects in the simulation, the vertical, dotted line shows in the initial mass given to embryos in each simulation, and the vertical, dash-dotted line shows the initial mass given to all planetesimals in each simulation. The color and shape of each point indicate the expansion factor used in the \reb\ simulation that produced that object.}
    \label{fig:uni_final_cmfs}
\end{figure}

Fig. \ref{fig:uni_final_cmfs} shows the masses and CMFs of the remaining objects from all 50 simulations along with the expansion factor used in the simulation that formed each object. Embryos from these simulations can form planets with a wide variety of masses, with the highest being nearly 30 times the size of the initial embryo mass. In addition to planets, the plot contains many fragments with masses below the planetesimal mass. A good number of these small objects have a mass of 0.0047$M_{\oplus}$, the minimum fragment mass and the lowest mass an object can have in the simulation. Finally, the plot has few objects that lie between the embryo and planetesimal mass. It's likely that smaller objects undergo disruptive collisions or get accreted by larger objects before they can reach masses within this range.

On the far left side of the plot, fragments and smaller objects show a wide distribution of final CMFs. Most of these small objects are concentrated around the starting CMF of 0.3; however, a number of them were able to achieve more extreme CMFs, with some reaching CMFs as high as Mercury's (0.7) and some reaching a CMF of 0.0. The CMFs of objects larger than the embryo mass don't vary much from the initial value of 0.3. While it appears that smaller planets tend to deviate more from this value than larger planets, most of them have a final CMF that lies well within a range of 0.25 and 0.35, far from the extreme values observed in Mercury and some exoplanets. However, we did produce an object above the embryo mass that has a CMF below 0.1, so reaching these extreme values with our model is possible. 

This object had a mass of approximately 0.244 $M_{\oplus}$ when -- near the end of its formation -- it experienced an erosive, head-on collision that removed about 62 percent of its mass and changed its CMF from 0.286 to 0.0. It underwent several more collisions after this one, bringing its final mass to approximately 0.106 $M_{\oplus}$, its final CMF to 0.037, its final semi-major axis to 0.471 AU, and its final eccentricity to 0.255. Effectively, this collision ejected the core of the planet and left it solely composed of mantle material. If users deem this collisional outcome to be too nonphysical, then $f\textsubscript{min}$ and $f\textsubscript{max}$ can be adjusted to avoid such a result.

The relationship between CMF and mass is most likely due to the averaging effects of accretion collisions and mergers. Smaller objects are often the products of disruptive collisions and are also more vulnerable to disruptive collisions \citep{Asphaug2010}, leading to a wide range of CMFs. Larger objects can experience these disruptive collisions as well but must also accrete a large amount of material to reach such high masses, which will tend to bring their CMFs closer towards the average value for the system. This process and the conditions of these simulations seem to indicate that extreme disruptive collisions late in a planet's formation are the most likely way giant impacts can produce the high CMFs seen in observations. In our data, a late erosive impact produced the only planet with an extreme CMF, and it is possible that Mercury and other iron-rich exoplanets were formed from a similar impact or impacts. These results could also indicate that larger planets with high CMFs formed from embryos that initially had high CMFs, a possibility that we discuss further in \S~\ref{sec:varying_disc_comp_final_cmfs}.

The treatment of collisional fragments in these simulations may also explain the lack of large CMF variations. If a disruptive collision produces fragments that have masses that are smaller than the minimum fragment mass, then this collision is instead considered a merger. Therefore, many collisions that could have been erosive or accretive are instead considered mergers in these simulations. In addition, because fragments from collisions often remain in the system, they can be reaccreted by the largest remnant, which will move its CMF closer towards its pre-collision value. Decreasing the minimum fragment mass may lead to more extreme results for planetary CMFs, though this will also increase the number of particles in a simulation, thereby increasing the simulation's computational time. Adding a debris loss prescription could create a more diverse distribution of CMFs \citep{Scora2022} though this will also lead to the creation of smaller planets.

Another possibility is that the disc is too dynamically cold to produce enough energetic collisions to create a larger variety in the CMFs of planets. Increasing the initial eccentricities and inclinations of the objects in the disc could lead to more energetic collisions, though re-accretion will still be a problem \citep{Scora2022}. Additionally, adding the migration of outer giant planets during the simulation can perturb objects in the disc and lead to more disruptive collisions \citep{Carter2015}.

\subsubsection{Model Data}
\label{sec:model_data}
To ensure the DBCT is properly implementing the model described in \S~\ref{sec:model}, we record the compositional outcomes produced by the code for all disruptive collisions from the \Nbody\ simulations described in \S~\ref{sec:nbody_simulations}. In particular, we will study how the ejecta CMF determined by the DBCT depends on the impact parameter of a particular collision, which is the main parameter used for the model in \S~\ref{sec:model}. Because we want to study collisions that apply to the DBCT, we remove all non-disruptive collisions from our data set. We also remove all hit-and-run collisions where the target does not lose or gain mass to focus on impacts where the target can obtain a new CMF.

\begin{figure}
	\includegraphics[width=\columnwidth]{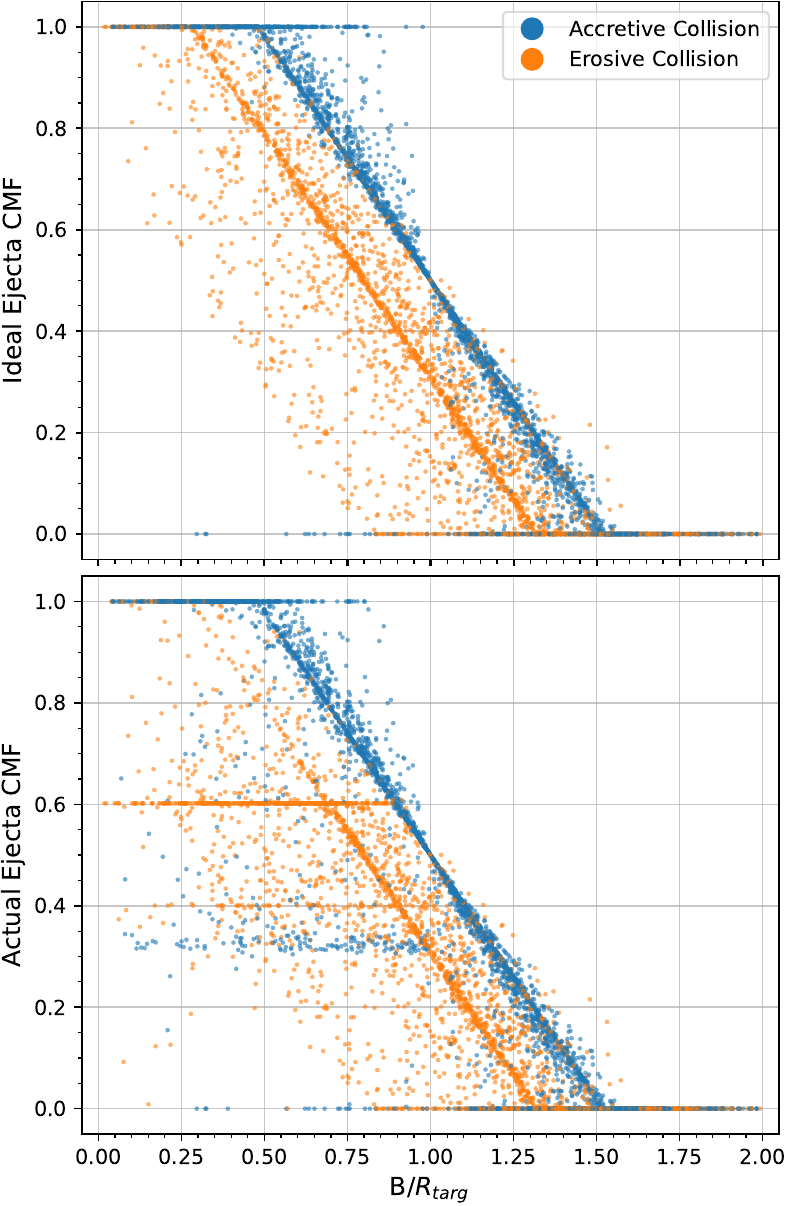}
    \caption{Plot showing how the CMF of collisional ejecta depends on the impact parameter normalized with respect to the outer radius of the target in the collision. The top plot shows the ideal ejecta CMFs that were calculated by our model while the bottom plot shows the CMFs that the ejecta actually obtained. This actual CMF can differ from the ideal one calculated by our model depending on the availability of core and mantle material to put into the ejecta after a disruptive collision. Blue points represent collisions where mass was transferred from the projectile to the ejecta, and orange points represent collisions where mass was transferred from the target to the ejecta.}
    \label{fig:B_vs_ideal_and_actual_CMF}
\end{figure}

The top panel of Fig. \ref{fig:B_vs_ideal_and_actual_CMF} shows how the the CMF of the collisional ejecta predicted by the model (which we name the ideal CMF) depends on the impact parameter with respect to the outer radius of the target. The graph has two prominent diagonal lines: one on the left containing mostly erosive collisions and one on the right containing mostly accretive collisions. As expected, the calculated CMF for the ejecta has an overall linear dependence with the impact parameter of the collision, an indication that the DBCT is applying the model properly.

The bottom panel of Fig. \ref{fig:B_vs_ideal_and_actual_CMF} shows how the actual CMF the ejecta obtains depends on the impact parameter with respect to the outer radius of the target. The actual CMF is determined by the availability of core material in the colliding bodies and can differ from the ideal CMF calculated by the model. In general, the plot contains more points scattered off its main two diagonal lines, which represent collisions where either the target or projectile did not have enough core or mantle material to match the ideal CMF that was calculated by our model. This panel also contains a horizontal line of erosive collisions at a CMF of 0.6 and a horizontal line of accretive collisions around a CMF of 0.3.

\begin{figure}
	\includegraphics[width=\columnwidth]{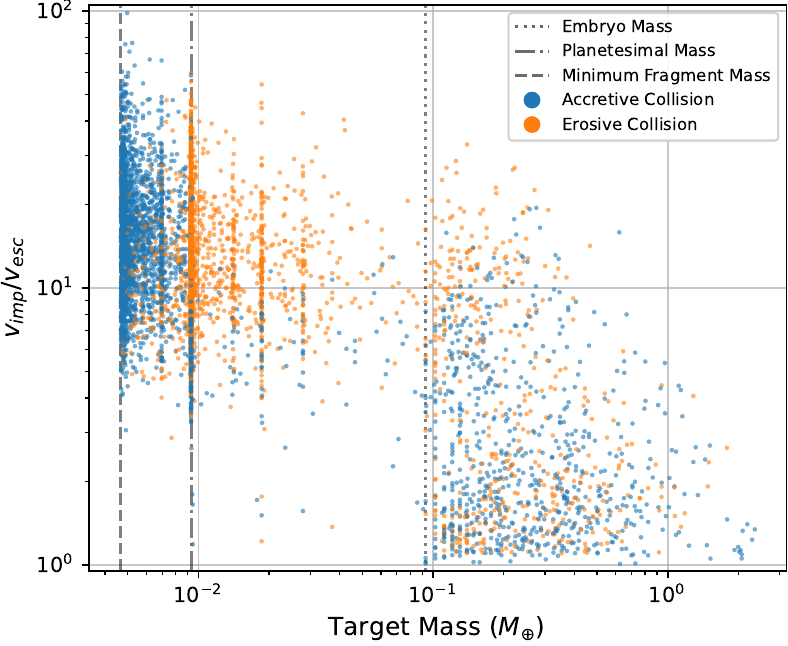}
    \caption{Plot showing the mass of the target in all disruptive collisions and the velocity of the impact with respect to the mutual escape velocity of the target and projectile. Blue points represent collisions where mass was transferred from the projectile to the ejecta, and orange points represent collisions where mass was transferred from the target to the ejecta. The dotted line shows the initial mass given to embryos in each simulation, the dash-dotted line shows the initial mass given to all planetesimals in each simulation, and the dashed line shows the minimum fragment mass for every simulation.}
    \label{fig:target_mass_v_imp}
\end{figure}

In order to find the origin of the horizontal lines in the bottom panel of Fig. \ref{fig:B_vs_ideal_and_actual_CMF}, we create Fig. \ref{fig:target_mass_v_imp}. The plot shows the target mass plotted against the impact velocity normalized with respect to the mutual escape velocity of the target and projectile for every disruptive collision across all 50 \reb\ simulations. The figure indicates that collisions involving high-mass targets often have a lower normalized impact velocity than collisions involving low-mass targets, which is likely due to the increase in the mutual escape velocity that occurs with higher target and projectile masses. The plot also indicates that objects with the mass of a planetesimal experienced a large number of erosive collisions while objects between the planetesimal and minimum fragment mass experienced a large number of accretive collisions.

These results offer a possible explanation for the horizontal lines in the bottom panel of Fig. \ref{fig:B_vs_ideal_and_actual_CMF}. The line of erosive collisions at a CMF of 0.6 could represent a specific outcome of collisions involving planetesimals in the initial disc that occurred multiple times throughout the simulations. In this type of collision, the planetesimal is the target of the erosive collision and produces ejecta that has half the mass of the planetesimal, which corresponds to the minimum fragment mass for these \reb\ simulations. Because the planetesimals all have an initial CMF of 0.3, the maximum CMF any material ejected from this object can obtain is 0.6. Even though our model calculates that the ejecta should obtain a CMF higher than 0.6, the planetesimal doesn't have enough core material to meet this calculation, and therefore, the code gives the ejecta the maximum possible value for its CMF. The line of accretive collisions at a CMF of 0.3 could have a similar explanation, though the exact conditions for these collisions may be harder to discern since an accretive collision takes core material from the projectile and not the target.

\subsubsection{Expansion Factor}
\label{sec:expansion_factor}
\citet{Childs2022} explored how the addition of an expansion factor to the objects in an \Nbody\ simulation affects the final architecture and other properties of the newly created planetary system. Here we explore how the expansion factor -- which changes the radii and densities of the bodies in these simulations -- affects the final CMFs of the resulting planets, using the same data used in \citet{Childs2022}. 

\begin{figure}
\includegraphics[width=\columnwidth]{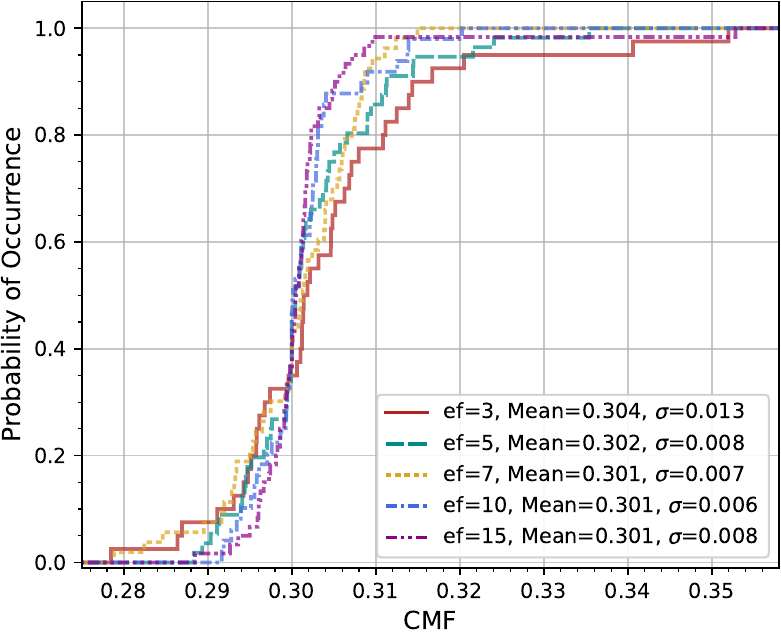}
    \caption{Plot showing the cumulative distributions for the CMFs of objects with masses greater than the initial embryo mass (0.093 $M_{\oplus}$) from all 50 fragmentation \reb\ simulations. The color and style of each line represents the expansion factor used for the 10 simulations that produced those planets. The plot also shows the average CMF and its standard deviation for each expansion factor.}
    \label{fig:ef_CDFs}
\end{figure}

\begin{table}
    \centering
    \caption{Statistics and P-values from Kolmogorov–Smirnov tests between the distributions shown for the different expansion factors in Fig. \ref{fig:ef_CDFs}}
    \begin{tabular}{cccc}
    \hline
    \hline
         expans. fact. (a) & expans. fact. (b) & Statistic & P-value\\
         \hline
         3 & 5 & 0.182 & 0.355\\
         3 & 7 & 0.164 & 0.490\\
         3 & 10 & 0.297 & 0.027\\
         3 & 15 & 0.281 & 0.033\\
         5 & 7 & 0.119 & 0.769\\
         5 & 10 & 0.181 & 0.298\\
         5 & 15 & 0.183 & 0.239\\
         7 & 10 & 0.252 & 0.057\\
         7 & 15 & 0.243 & 0.058\\
         10 & 15 & 0.150 & 0.504\\
         \hline
    \end{tabular}
    \label{tab:KS_test_table}
\end{table}
Fig. \ref{fig:ef_CDFs} contains cumulative distributions created using the results from \S~\ref{sec:uniform_disc_final_cmfs} for objects with masses greater than 0.093$M_{\oplus}$ (the initial embryo mass). The colors and styles of the lines in the plot denote the different expansion factors used in each simulation. We perform Kolmogorov–Smirnov tests between each of these samples to see if they came from the same population of final planetary CMFs; these tests will indicate if expansion factors affect the range of planetary CMFs produced by a simulation. The results of these tests are shown in Table \ref{tab:KS_test_table} and indicate that using different expansion factors can change the distribution of planetary CMFs in the final system produced by the fragmentation module and our DBCT. This effect is most prominent when comparing the planetary CMFs that resulted from the use of a low expansion factor (such as 3 or 5) and ones that resulted from the use of high expansion factors (such as 10 and 15). 

Our results imply that simulations using low expansion factors are more likely to produce planets with a wide range of CMFs.\footnote{The object from Fig. \ref{fig:uni_final_cmfs} with a CMF lower than 0.1 is not shown in Fig. \ref{fig:ef_CDFs} to prevent skewing. This object was produced by a simulation with an expansion factor of 3.} This result seems to corroborate findings from \citet{Childs2022}. In that study, the authors found that a lower expansion factor leads to a higher percentage of disruptive collisions compared to mergers; it's these disruptive collisions that transfer core and mantle material between objects and lead to CMF changes. Users must decide if a higher expansion factor is worth the decrease in disruptive collisions that will occur in their simulations.

\subsubsection{Minimum and Maximum CMF}
\label{sec:min_and_max_cmf}

\begin{figure}
\includegraphics[width=\columnwidth]{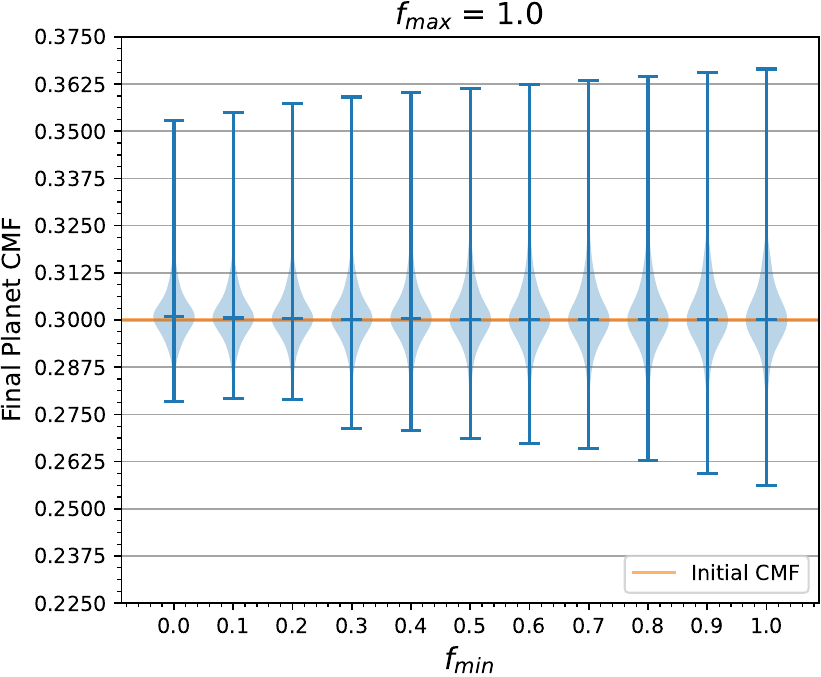}
    \caption{Violin plot showing how the final CMFs of planets depend on $f\textsubscript{min}$, the minimum fraction of core material that collisional ejecta can have in our DBCT. For this plot, we keep $f\textsubscript{max}$ at 1.0 while using ten different values of $f\textsubscript{min}$. The bars in the center of each line show the median of the final planetary CMFs for all 50 simulations. The bars at the end of each line indicate the maximum and minimum CMFs that were produced, and the orange line indicates the initial CMF for all objects in the disc.}
    \label{fig:changing_fmin}
\end{figure}

\begin{figure}
\includegraphics[width=\columnwidth]{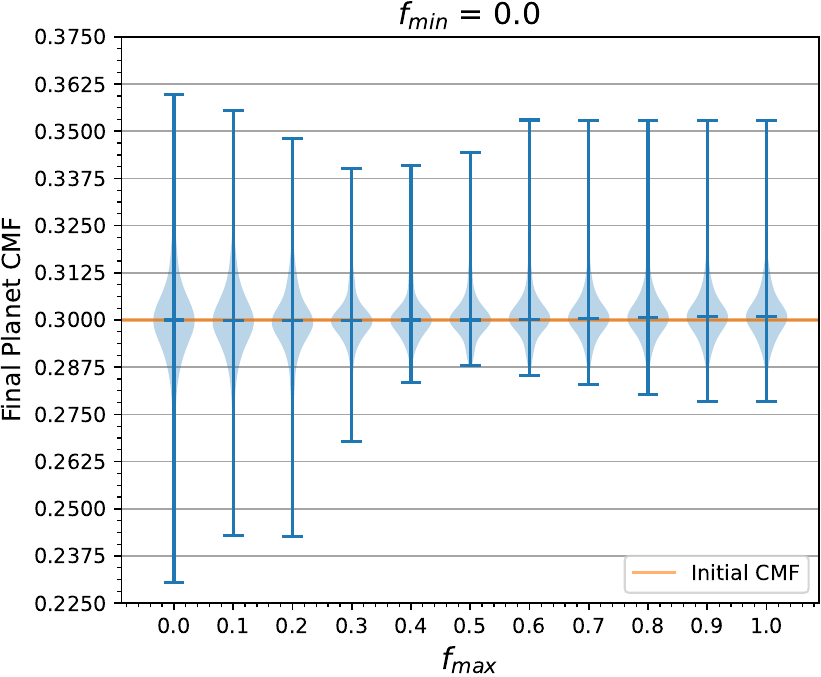}
    \caption{Same as Fig. \ref{fig:changing_fmin}, but $f\textsubscript{max}$ is varied while $f\textsubscript{min}$ is kept constant at 0.0.}
    \label{fig:changing_fmax}
\end{figure}

In our model, the minimum and maximum fractions of core material that compose the ejecta of disruptive collisions ($f\textsubscript{min}$ and $f\textsubscript{max}$) are variables that can be set by the user. They can have any value from 0.0 to 1.0, but $f\textsubscript{max}$ must always be greater than or equal to $f\textsubscript{min}$. To see how these variables affect the final CMFs of objects larger than the embryo mass, we run all 50 simulations through our DBCT with different combinations for $f\textsubscript{min}$ and $f\textsubscript{max}$.  Fig. \ref{fig:changing_fmin} and Fig. \ref{fig:changing_fmax} show the results of changing these two variables in our code. In Fig. \ref{fig:changing_fmin}, we vary $f\textsubscript{min}$ while keeping $f\textsubscript{max}$ constant at 1.0, and in Fig. \ref{fig:changing_fmax}, we vary $f\textsubscript{max}$ while keeping $f\textsubscript{min}$ constant at 0.0.

In both figures, we remove the object that produced the most extreme CMF for a planet-sized object in Fig. \ref{fig:uni_final_cmfs} whose collisional history was discussed in \S~\ref{sec:uniform_disc_final_cmfs}. We consider this object an outlier in our data since it has a CMF that deviates from the initial CMF three times more than any other planet. Therefore, we remove it in order to get a more accurate sense of how $f\textsubscript{min}$ and $f\textsubscript{max}$ affects the majority of planet-sized objects from our results. \S~\ref{sec:varying_disc_comp_final_cmfs} shows how much the CMF of this one planet can vary from using different model parameters. 

For both Fig. \ref{fig:changing_fmin} and Fig. \ref{fig:changing_fmax}, the distributions of planetary CMFs and their medians are centered around 0.3 -- the initial CMF for objects in the disc -- and don't appear to have any strong dependence on these two model parameters. However, both plots contain some variation in the maximum and minimum CMFs that planets can achieve with these different parameters, though no object in this data set reaches a CMF that is above 0.4 or below 0.2 (excluding the removed object). These results could change for collisional data that includes more extreme disruptive collisions involving planet-sized objects.

\subsection{Varying Disc Composition}
\label{sec:varying_disc_composition}

In \S~\ref{sec:uniform_disc}, we assume that all of the objects in our disc have the same initial CMF, which is often done in these types of studies \citep{Carter2015, Scora2020, Scora2022}. While this is a valid assumption, both \citet{Aguichine2020} and \citet{Johansen2022} posited ways that planetesimals in the inner part of the disc could become enriched in core material. In addition, \citet{Carter2015} and \citet{Scora2020} showed that the collisions that formed planetary embryos could give them a CMF deviation of approximately 10\% from the primordial composition of objects in the disc going into the giant impact phase of formation. Multiple studies \citep{Johansen2022, Scora2022, Mah2023} suggest that giant impacts between these core-enriched objects could have produced the high CMFs we observe in Mercury and some exoplanets. 

According to \citet{Johansen2022}, these iron-rich planetesimals formed from the accretion of iron-rich pebbles that originated near the iron evaporation line in the early planetary disc. Due to changes in stellar luminosity and other heating sources, the temperature of the disc will vary with time, which will change the position of the evaporation line in the disc and create a gradient of iron-rich pebbles throughout the disc instead of just a narrow region of them \citep{Mah2023}. Therefore, in this section, we use different distributions of core material in the initial disc described in \S~\ref{sec:nbody_simulations} and see how they affect the CMFs of planets from the data in \S~\ref{sec:uniform_disc_final_cmfs}.

\subsubsection{Initial Disc Distributions}
\label{sec:initial_disc_distributions}

We choose three different types of non-uniform distributions of core material that depend on an object's location in the disc to use with our DBCT: a step function distribution that contains three different regions of planetoid CMFs (0.1, 0.3, and 0.5); a linear distribution where an object at the inner edge of the disc has a CMF of 0.5 and an object at the outer edge has a CMF of 0.1; and an exponential distribution where an object at the inner edge of the disc has a CMF of 0.7 and an object at the outer edge has a CMF of 0.1. In all three, we assume that objects in the inner part of the disc will have a higher CMF than objects in the outer part of the disc and that these distributions will only depend on an object's semi-major axis. In addition, we create these distributions in such a way that the average initial CMF is around 0.3. The functions that describe these distributions may not be entirely physical, but for an initial investigation into this topic, they are a straightforward way to explore the consequences of a non-uniform disc. 

\begin{figure*}
\includegraphics[width=\textwidth]{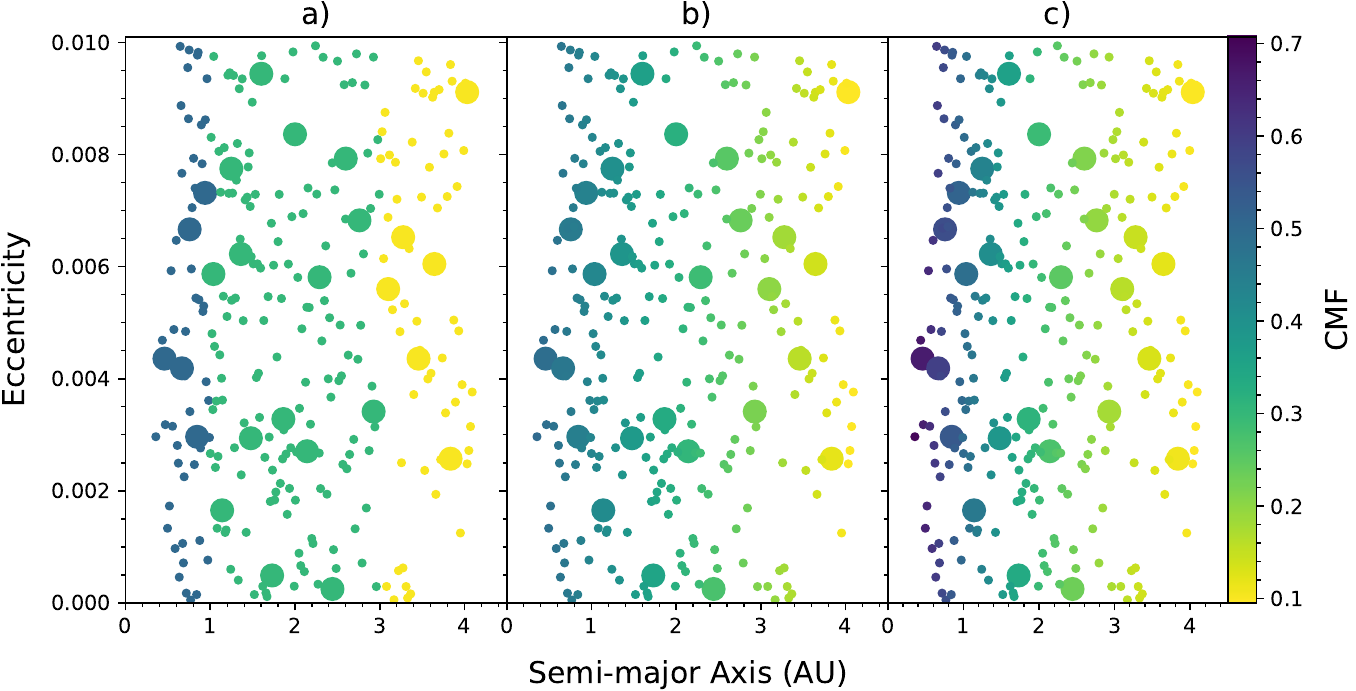}
    \caption{Plot showing the orbital parameters and compositions of an initial disc from one \reb\ simulation with three different distributions of core material. a) shows the step-function distribution with a CMF range of 0.1 to 0.5, b) shows the linear distribution with a CMF range of 0.1 to 0.5, and c) shows the exponential distribution with a CMF range of 0.1 to 0.7.}
    \label{fig:initial_distributions}
\end{figure*}

Fig. \ref{fig:initial_distributions} shows all three distributions applied to one of the initial discs from the simulations in \S~\ref{sec:nbody_simulations}. The leftmost panel shows the step-function distribution of CMFs where objects in the inner region of the disc have a CMF of 0.5, objects near the outer region of the disc have a CMF of 0.1, and objects in between these regions have a CMF of 0.3. This distribution gives the bodies in the initial disc a starting average CMF of 0.294 and can be described mathematically by
\begin{equation}
\label{eq:sf_distribution}
    F\textsubscript{obj} =  \begin{cases}
        0.5, &  a \leq 1.0\\
        0.3, &  1.0 < a < 3.0\\
        0.1, &  a \geq 4.0
    \end{cases},
\end{equation}
where $F\textsubscript{core}$ is the CMF of an object in the disc and $a$ is the object's initial semi-major axis. 

The middle panel of Fig. \ref{fig:initial_distributions} shows the linear distribution of CMFs where objects at the inner edge of the disc have a CMF of 0.5 and objects at the outer edge of the disc have a CMF of 0.1. The starting CMF of a particular planetoid will linearly decrease the further its position is from the inner edge of the disc. This distribution gives the bodies in the initial disc a starting average CMF of 0.314 and was created with the following equation:
\begin{equation}
F\textsubscript{obj} = -0.1096(a-0.35)+0.5.
\label{eq:lin_distribution}    
\end{equation}
Our innermost body begins at $0.35 \, \rm au$, so we subtract 0.35 from all $a$ values to ensure all CMFs are between 0.1 and 0.5.

The right panel of Fig. \ref{fig:initial_distributions} shows the exponential distribution of CMFs where objects can have a maximum CMF of 0.7 and a minimum CMF of 0.1. The starting CMF of a particular planetoid will exponentially decrease the further its position is from the inner edge of the disc. This distribution gives the bodies in the initial disc a starting average CMF of 0.326 and uses the equation:
\begin{equation}
F\textsubscript{obj} = 0.7e^{-0.5331(a-0.35)}.
\label{eq:exp_distribution}    
\end{equation}

We use all three CMF distributions to create compositional input files for all 50 simulations. We then use our DBCT to see how these distributions affect the final CMFs of planets from all simulations. For each distribution and simulation, we set $f\textsubscript{min}$ and $f\textsubscript{max}$ to 0.0 to maximize the amount of mantle material transferred during each collision, which will create the ideal conditions to produce planets with high CMFs. 

\subsubsection{Final CMFs}
\label{sec:varying_disc_comp_final_cmfs}

\begin{figure*}
\includegraphics[width=\textwidth]{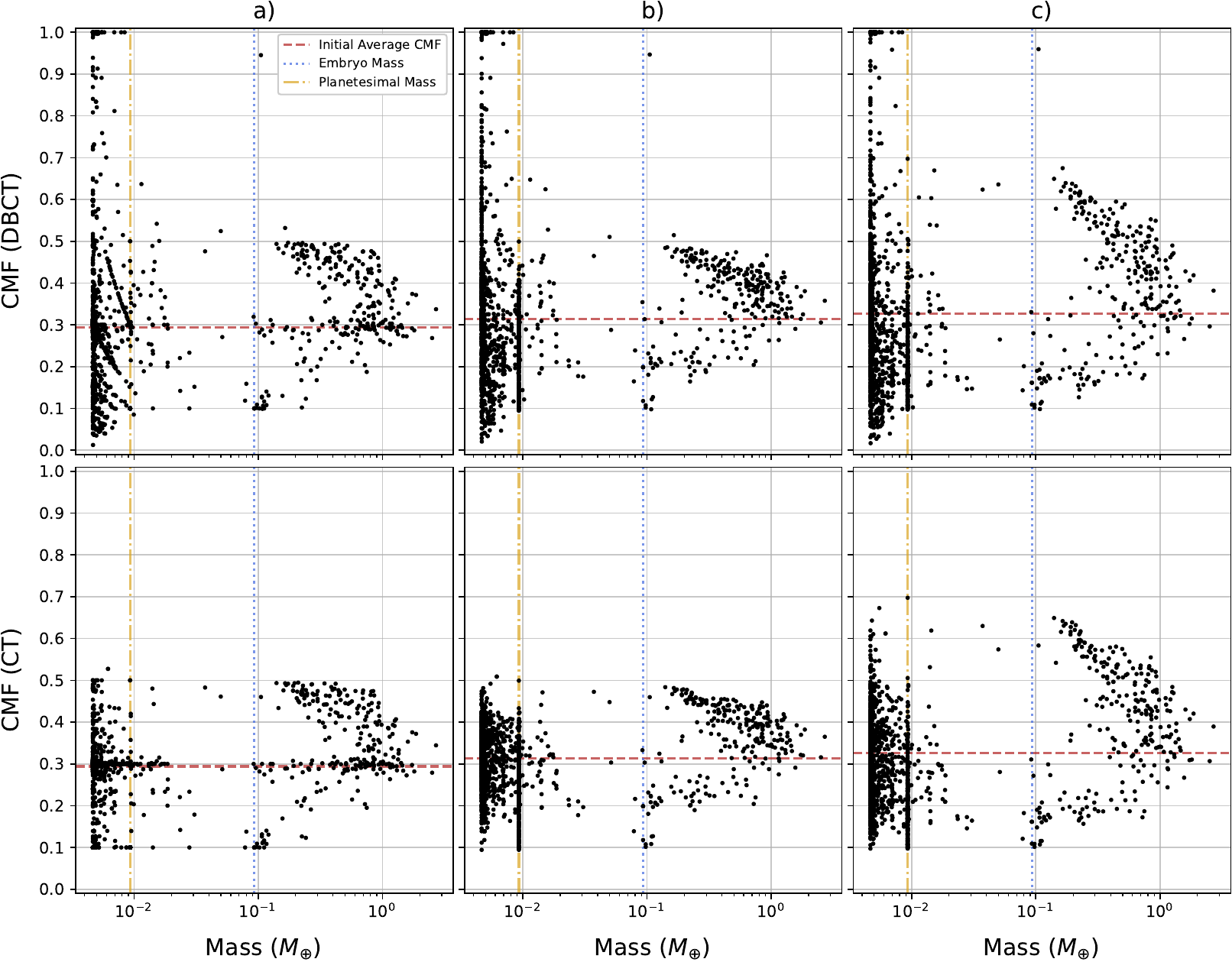}
    \caption{Final Masses and CMFs for the remaining objects in all 50 \reb\ simulations with different distributions of core material in the initial disc. a) shows the step-function distribution, b) shows the linear distribution, and c) shows the exponential distribution. The top row shows results from our DBCT (DBCT) and the bottom row shows results from the composition tracking code (CT) created by \citet{Childs2022}. The blue, dotted line shows the initial mass given to embryos in each simulation, the orange, dash-dotted line shows the initial mass given to all planetesimals in each simulation, and the red, dashed line shows the average CMF of the initial disc for each distribution.}
    \label{fig:three_final_cmfs}
\end{figure*}

The top row of Fig. \ref{fig:three_final_cmfs} shows the final CMFs of all remaining objects from all 50 \reb\ simulations after being processed by our DBCT. The leftmost panel shows the results from the step function distribution, the middle panel shows the results from the linear distribution, and the rightmost panel shows the results from the exponential distribution. These results demonstrate the same relationship between mass and CMF that can be seen in \ref{sec:uniform_disc_final_cmfs}: as objects collect more mass, their CMF moves towards the average value for the core material that remains in the system. This relationship produces the arrowhead shape in the plots for objects with masses greater than the embryo mass and remains consistent even for initial average CMFs that are not around 0.3 \footnote{We have verified this claim using a step-function distribution that has an initial average of 0.494}.

All three plots in the top row of Fig. \ref{fig:three_final_cmfs} have more planets with CMFs that are above the average initial value for that distribution -- indicating that they have been enriched with core material -- and a large number of collisional fragments with CMFs that are below this average -- indicating that they have been enriched with mantle material. This result could be due to the values of $f\textsubscript{min}$ and $f\textsubscript{max}$ that were chosen to maximize the amount of mantle material in the collisional ejecta for objects in these distributions. However, this result could also be due to ejections of the initially low-CMF bodies in the outer disc due to their interactions with Jupiter and Saturn.\footnote{In these \reb\ simulations, once a particle's semi-major axis reaches 100 AU, it is considered ejected from the system and removed from the simulation.} After further investigation using initial discs with core-rich outer regions, we find that these ejections are the most likely explanation for the discrepency between high-CMF and low-CMF planets and fragments.

\begin{table}
    \centering
    \caption{Post-simulation compositional statistics for objects with masses greater than 0.93$M_{\oplus}$ from all three initial distributions.}
    \begin{tabular}{lccc}
    \hline
    \hline
         Distribution & Average CMF & Std. Dev. & Median CMF \\
         \hline
         Step-Function & 0.336 & $\pm 0.082$ & 0.368\\
         Linear & 0.365 & $\pm 0.096$ & 0.387\\
         Exponential & 0.403 & $\pm 0.147$ & 0.423\\
         \hline
    \end{tabular}
    \label{tab:planet_stats}
\end{table}

Table \ref{tab:planet_stats} shows the final compositional statistics for objects with masses greater than 0.093$M_{\oplus}$ (which we consider to be planets) in Fig. \ref{fig:three_final_cmfs} for all three of our initial distributions. For all distributions, the final average CMF of planets is higher than the average CMF for objects in the initial disc, which is likely due to the previously mentioned mantle stripping collisions and ejections of mantle-rich bodies.

Using these statistics, we approximate the mass at which a planet will have a CMF that starts to converge towards the final average planetary CMF in order to distinguish the types of objects that are more likely to have extreme CMFs. To quantify this cutoff mass for each distribution, we first obtain the masses of planets that have a CMF that's at least one standard deviation away from the final average planetary CMF. We then calculate the mean for these planetary masses to obtain the cutoff mass for that distribution. We find that for the step-function distribution, this mass cutoff is approximately 0.296$M_{\oplus}$; for the linear distribution, this mass cutoff is approximately 0.279$M_{\oplus}$; and for the exponential distribution, this mass cutoff is approximately 0.287$M_{\oplus}$. The step-function distribution has 39 planets beneath its cutoff mass, the linear distribution has 50, and the exponential distribution has 66.

All three distributions produced a single object above the embryo mass with a CMF that is greater than the CMF of Mercury (0.7), indicating that a high-CMF planet formed by giant impacts is possible but perhaps uncommon. This high-CMF planet produced by all three distributions is the same one discussed in \S~\ref{sec:uniform_disc_final_cmfs} that experienced a strong erosive collision near the end of its formation that drastically changed its final CMF. The embryo of this planet started near the inner edge of the disc and was therefore enriched with core material from the beginning of its formation, which also partly explains the high CMF it obtained in each distribution. Additionally and perhaps more significantly, the values for $f\textsubscript{min}$ and $f\textsubscript{max}$ differ between \S~\ref{sec:uniform_disc} and this section, leading to different outcomes for the planet's final composition after the erosive collision.

The formation of this high-CMF planet is one way to explain the compositions of Mercury and similar exoplanets: a core-enriched embryo accretes other core-enriched bodies and then experiences one or multiple erosive collisions that increase its CMF to a value that's similar to Mercury's. Arguably, our simulated planet obtained a CMF that's not comparable to Mercury's ($>$0.9 in all three distributions); however, the parameters we used for our model in this section assume disruptive collisions only transfer mantle material, which means the CMFs we obtained should be considered upper-bounds for this planet. The addition of SPH calculations into our model could help refine results such as this one.

The exponential distribution (top right panel of Fig. \ref{fig:three_final_cmfs}) produced multiple planets that have CMFs between 0.6 and 0.7, which is in the range of CMFs observed for some super-Mercuries. These planets likely began as core-enriched embryos near the inner edge of the disc with initial CMFs that were already between 0.6 and 0.7. Over the course of their formation, they did not accrete a large amount of mass, which would have lowered their CMFs towards the system average, and did not experience any extreme erosive collisions that could have increased their CMFs beyond this range. 

These results imply that Mercury and other exoplanets could have obtained their current CMFs as embryos before the giant impact phase of formation and then retained this CMF after experiencing relatively few collisions or only colliding and merging with other objects that have high CMFs. This pathway for formation would require the disc to have enough concentrated core material for planetary embryos with CMFs as high as 0.7 to form. Using estimated conditions of the early Solar System, \citet{Mah2023} were able to produce pebbles with CMFs as high as 0.6 in regions around evaporation lines in an early planetary disc and predicted that lower stellar metallicties could produce pebble CMFs that are higher. It's possible that these pebbles could accrete into core-enriched planetesimals and then core-enriched embryos that have CMFs as high as the ones seen in Mercury and other planets.

\subsubsection{Structure in the Step Function Distribution}
\label{sec:structure_in_step_function_distribution}

In Fig. \ref{fig:three_final_cmfs}, the top left panel shows the final CMFs of objects from the three-tiered step-function distribution of core material that were processed using our DBCT. In the region to the left of the planetesimal mass line, the plot contains three lines which originate at the intersection between the planetesimal mass and one of the starting CMF values for the distribution (0.1, 0.3, and 0.5). The lines have a negative slope that becomes much steeper the higher the starting CMF becomes.

We posit that these lines are a consequence of the parameters we used with the DBCT to make these figures and also due to the nature of the step function distribution itself. Both $f\textsubscript{min}$ and $f\textsubscript{max}$ were set to 0.0, which means the ejecta from a disruptive collision always receive a CMF of 0.0 if there is enough material from the target or projectile to do so. If an object only undergoes disruptive collisions and no mergers, then its final CMF can be predictable depending on the CMF it starts with due to the nature of the collisions in the \reb\ fragmentation module. Therefore, the initial planetesimals in these step-function distributions that don't undergo any mergers will have final CMFs that lie along a predictable evolutionary path, which appear as these lines in the top left panel of Fig. \ref{fig:three_final_cmfs}.

\begin{figure}
\includegraphics[width=\columnwidth]{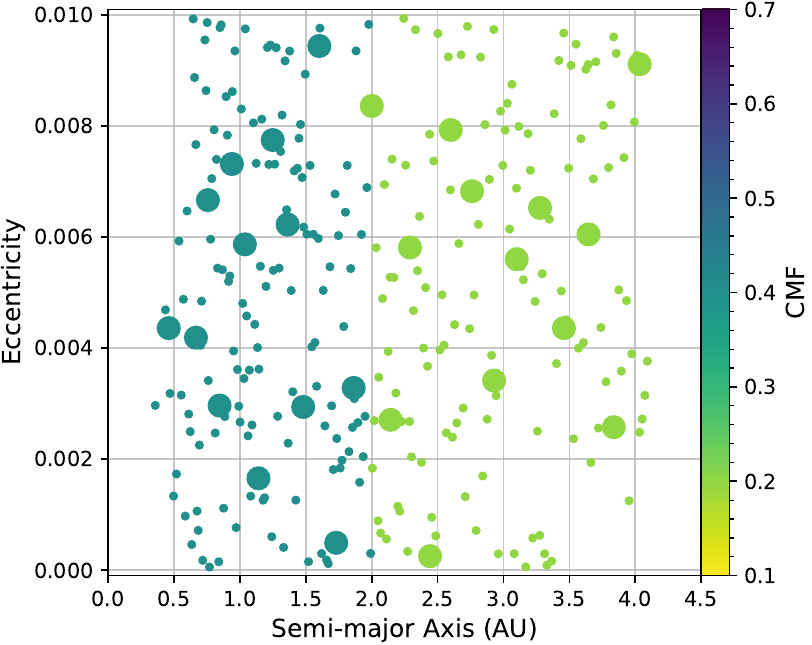}
    \caption{Plot showing the orbital parameters and compositions of an initial disc from one \reb\ simulation with a step function distribution of two possible starting CMFs.}
    \label{fig:2step_distribution}
\end{figure}

We will attempt to confirm that these three lines are a result of the DBCT model parameters and not some other phenomenon. We create a new, simplified step function distribution for the initial CMFs of objects in our disc. Instead of containing three different initial CMFs, this distribution will contain only two possible initial CMFs. Objects from the inner edge of the disc up to 2.0 AU will have a CMF of 0.4 and objects with a semi-major axis greater than 2.0 AU will have a CMF of 0.2. With these conditions, the disc for this distribution has an initial average CMF of 0.304. Fig. \ref{fig:2step_distribution} shows this disc, which can be described with the following equation: 
\begin{equation}
\label{eq:2step_distribution}
    F\textsubscript{obj} =  \begin{cases}
        0.4, &  a \leq 2.0\\
        0.2, &  a > 2.0
    \end{cases},
\end{equation}

\begin{figure}
\includegraphics[width=\columnwidth]{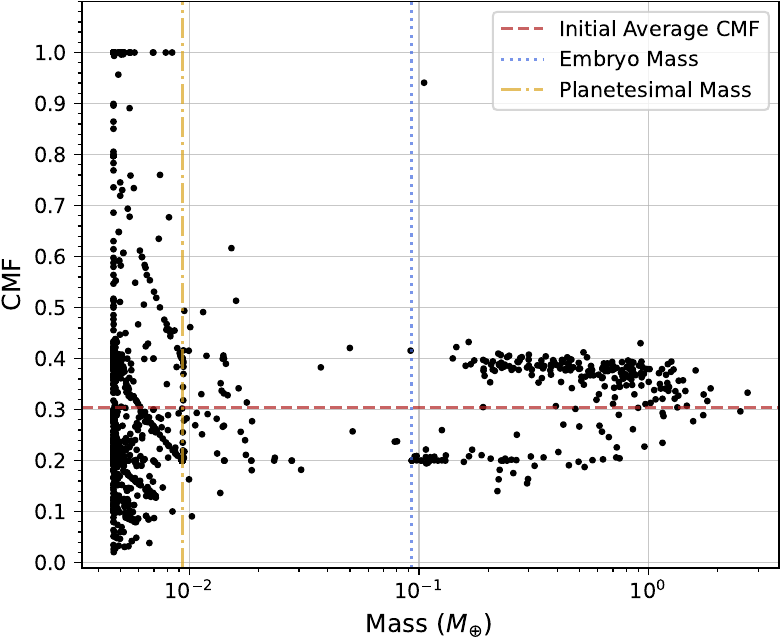}
    \caption{Final Masses and CMFs for all remaining objects from the 50 simulations with the step function distribution shown in Fig. \ref{fig:2step_distribution} applied to each initial disc. The red, dashed line shows the average CMF of the initial disc for each distribution, the blue, dotted line shows the initial mass given to embryos in each simulation, and the orange, dash-dotted line shows the initial mass given to all planetesimals in each simulation.}
    \label{fig:2step_final_cmfs}
\end{figure}

Fig. \ref{fig:2step_final_cmfs} shows the results of using this bimodal distribution of core material for all 50 \reb\ simulations with the same model parameters that were used in \S~\ref{sec:varying_disc_composition}. The results once again show a wider variety of CMFs for objects below the planetesimal mass than for objects above the embryo mass and that larger planet-sized objects tend to have CMFs that are closer to the average for the system. In addition, the plot contains more objects with CMFs above the average than below the average, which is due to a combination of the model parameters and the ejection of core-depleted material in the outer disc.

The figure contains two lines beginning at the planetesimal mass and CMFs of 0.2 and 0.4, which correspond to the starting CMFs in this two-tiered distribution. This result indicates that the the number of lines in these plots is dependant on the number of tiers in a step-function distribution. We again argue that these lines are simply a result from the parameters we have chosen to use with our DBCT for these distributions and do not come from some sort of physical process.

Though our two-step distribution only applies to the CMF of an object, \citet{Childs2022} performed a similar study looking at how the water mass fractions (WMFs) of the remaining objects in a planetary system depend on the initial distribution of water in the disc. Both of our distributions are bimodal and depend solely on an object's starting position in the disc with the densest objects (iron-rich and water-poor) starting close to the inner edge of the disc and the least dense objects (iron-poor and water-rich) starting further out.  At this time, we do not consider any vaporization of the planetary materials, which may affect planet composition, especially when the material is volatile.

This step-function composition could be reflected in the planet-forming materials due to the material being on either side of the water ice line.  The fact that this composition profile yields two composition branches that only converge at the largest masses might provide an explanation for the observation that there are two primary types of terrestrial exoplanets---water rich and water poor\citep{Luque:2022}.  In this scenario, terrestrial planet formation didn't persist long enough to merge the material from the water-rich and water-poor parts of the initial disc.

\subsubsection{Final Semi-Major Axes}
\label{sec:final_semi_major_axes}

\begin{figure*}
\includegraphics[width=\textwidth]{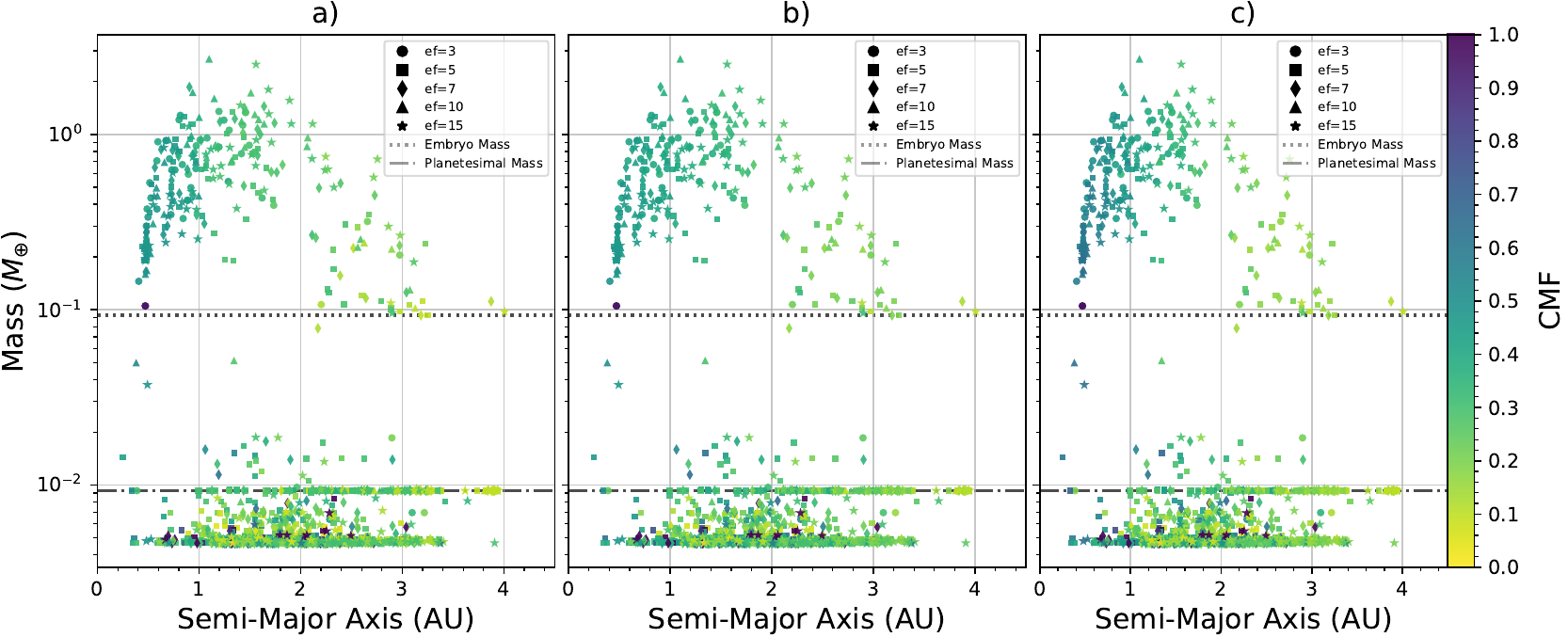}
    \caption{Plot showing the semi-major axes, masses, and compositions of all remaining objects from all 50 \reb\ simulations with different distributions of core material in the initial disc.  a) shows the step-function distribution, b) shows the linear distribution, and c) shows the exponential distribution. The dashed line shows the initial mass given to embryos in each simulation, and the dash-dotted line shows the initial mass given to all planetesimals in each simulation. The colors of each point represent the CMFs of these objects and range from values of 0.0 to 1.0. The shape of each point indicates the expansion factor used in the \reb\ simulation that produced that object.}
    \label{fig:semi-major_axis_plots}
\end{figure*}

In order to determine the amount of radial mixing that occurred in the disc during these simulations, we extract the semi-major axes for the objects that remain at the completion of each simulation. To collect this data, we utilize the simulation archive feature in \reb\ that allows users to save the orbital and physical parameters of objects in a simulation at a particular simulation time; users can then use these archives to restart simulations and analyze this data.

Fig. \ref{fig:semi-major_axis_plots} shows the semi-major axes, masses, and compositions for all remaining objects from all 50 \reb\ simulations. The different panels indicate which initial distribution of core material was used during the analysis with the DBCT: panel a) is the step-function distribution, panel b) is the linear distribution, and c) is the exponential distribution. The axes represent the semi-major axis and mass of each object, which remain the same between panels. The only feature of the panels that differ are the colors of the points, which represent the CMF of each body.

Upon inspection, the figure contains no discernible trends between mass and semi-major for objects with masses below the fragment mass. However, for objects with masses above the embryo mass, the graph shows two relationships between these parameters. First, the graph contains less planet-sized objects at higher semi-major axes, especially past 2.0 au. This is likely the result of two separate factors: the negative slope of the surface density profile in the initial disc and the higher number of ejections for objects further out in the disc. Second, the masses of planets tend to increase as the semi-major axis moves closer to the center region of the initial disc (around 2.0 au), indicating that embryos near the center of the disc are able to accrete material more effectively then embryos towards the edges of the disc. 

The CMFs of bodies below the fragment mass do not appear to have any relationship with semi-major axis. However, planet-sized objects show a clear trend between CMF and semi-major axis in all plots. In all three of these panels, the final distributions of these planetary CMFs seem to generally reflect the type of distribution that was implemented in the initial disc (step-function, linear, and exponential). This result indicates that any radial mixing and erosive collisions that occured were not enough to drastically change the original distribution of core material across the system. The spread of core material in the initial disc will likely indicate how this material will be distributed in the resulting planets. 

\subsection{Comparison to Composition Tracking Code}
\label{sec:comparison_to_comp_tracking}

To see how our DBCT compares to the composition tracking code created by \citet{Childs2022} -- which assumes that the core and mantle material in an object are uniformly mixed -- we enter all three previously described initial distributions and all 50 \reb\ simulations into the composition tracking code and see how the final compositions of the remaining objects compare to those found with the DBCT\footnote{CMF for the composition tracking code refers to the fraction of what's considered core material in the object's homogeneous composition.}. These results have been plotted in the bottom row of Fig. \ref{fig:three_final_cmfs} and show some of the differences between the two codes.

Most noticeably, the DBCT has a higher CMF variance than the composition tracking code, especially for smaller bodies. Additionally, the composition tracking code does not produce the planet with a CMF greater than 0.9 that can be seen in the results for the DBCT. Because the composition tracking code assumes that objects are undifferentiated, mantle and core material cannot be preferentially transferred in a collision. This prevents any extreme CMFs from being created for all objects in a simulation, including the smaller objects that are most vulnerable to disruptive collisions. This finding highlights why differentiation should be taken into account when performing studies involving disruptive collisions in \Nbody\ simulations, particularly ones investigating the formation of high-CMF objects. 

\section{Discussion}
\label{sec:Discussion}
Our study and previous \Nbody\ studies have encountered difficulties in consistently producing planets with high CMFs, especially ones with large masses. Even with favorable conditions for disruptive collisions that remove mantle material, \citet{Scora2022} were only able to produce one planet with a CMF around 0.6\footnote{Each object in their initial disc began with a CMF of 0.33}. \citet{Cambioni2021} were able to produce a few high CMF objects, including one with a similar composition and mass to Mercury, but did not achieve a CMF higher than 0.6 for objects with a mass greater than 0.1$M_{\oplus}$. It's possible that the scarcity of high-CMF planets produced by these studies reflects the actual abundance of planets with high CMFs; the large percentage of iron-rich exoplanets that have been observed could be a result of observation bias towards these types of dense planets \citep{Burt2018}. However, core-enriched super-Mercuries have been observed with masses much greater than that of Earth, something no \Nbody\ study has been able to replicate. Future work should attempt to explain how these planets could grow to these sizes with such high CMFs.

When it comes to giant impacts, our work and previous work by \citep{Scora2022} suggest that disruptive collisions near the end of a planet's formation are the most likely way to produce a high planetary CMF. An embryo could go through energetic collisions early on in its formation, but if it accretes a large amount of mass afterwards -- which objects tend to do if a disruptive collision happens early in their formation \citep{Quintana2016} -- then its CMF will tend to move back towards the average CMF of the system. Migration of giant planets could be a possible source of perturbation that could create these necessary collisions, but its unclear if such a process occurs in a large number of planetary systems. However, future work that includes a steady source of perturbations throughout the formation process could be used to see how they affect final planetary compositions.

Additionally, a closer look at how debris-loss and fragments from a collision are treated would be informative. During collisions, it's possible that a large amount of material is turned into particles small enough to be removed from the system by radiative forces on a short time-scale, which will prevent re-accretion from occurring \citep{Jackson2012, Scora2022}. In addition, it may be that larger variations in the CMFs of objects in the initial disc could be an important factor in the formation of high CMF planets. Overall, it's plausible that the extreme CMFs that have been observed in Super-Mercuries are due to a combination of all of the previously mentioned effects, many of which have been individually explored in previous studies. It will be insightful to try and perform simulations with different groups of them active at once.

Performing more SPH calculations will improve the detail of disruptive collision outcomes and how material is transferred between objects during these impacts. These new results can produce updated models for tracking composition changes that can be added to \Nbody\ simulations, giving future studies of planet formation better predictions on how the CMFs of objects will change for collisions with a variety of parameters. It will also be beneficial to calculate the transfer of core and mass material during collisions in \Nbody\ simulations instead of performing these calculations afterwards in a post-processing code. Mantle stripping collisions can change the masses, radii, and other physical properties of objects, which could affect the evolution of a planetary system and the distribution of core and mantle material over time.

Future observations of extrasolar planets will also be important in trying to narrow down the ways that these high and low CMF planets form. As previously mentioned, it's possible that the percentage of iron-enriched planets that have been discovered is simply a result of observation bias \citep{Burt2018} and these types of planets rarely occur in most systems. Additional data will need to be collected to determine whether compositions that vary from stellar metallicity are a normal feature in planetary systems.

\section{Conclusion}
\label{sec:conclusion}

In this paper, we presented a post-processing code called the Differentiated Body Composition Tracker (DBCT) that allows users to track how core and mantle material is transferred between differentiated objects after disruptive collisions. The model implemented in this code primarily depends on the impact parameter of a collision and allows users to set the minimum and maximum fraction of core material that can comprise the collisional ejecta ($f\textsubscript{min}$ and $f\textsubscript{max}$). This code is designed to work with collisional data that is collected from \reb\ simulations with the fragmentation module for collisions enabled \citep{Childs2022}. For all DBCT results in this manuscript, we used data collected by \citet{Childs2022} from 50 \reb\ simulations containing a disc with 26 mars-sized embryos, 260 moon-sized planetesimals, Jupiter, and Saturn.

We first assumed that all objects in the initial disc began with a uniform CMF of 0.3 and set $f\textsubscript{min}$ = 0.0 and $f\textsubscript{max}$ = 1.0, the widest possible range between these two model parameters, before entering the collisional data from \reb\ into the DBCT. We found that collisional fragments produced the most varied compositions, with some fragments being composed of all mantle (CMF = 0.0) and some being composed of all core (CMF = 1.0). The majority of objects above the embryo mass -- which we considered to be planets -- remained close to their initial compositions, with only one planet containing a CMF farther than 5\%\ from its initial value of 0.3. This particular planet experienced an erosive, head-on collision that removed about 62 percent of its mass near the end of its formation and ended the simulation with a CMF of 0.037.

We then varied $f\textsubscript{min}$ and $f\textsubscript{max}$ to better understand how these parameters in the DBCT affect the final CMFs of objects. We found that the distribution of planetary CMFs from all 50 simulations does not have a strong dependence on these two parameters, though this result could possibly change if the simulation data contains more disruptive collisions. We also explored how the inclusion of an expansion factor in \Nbody\ simulations can affect the final compositions of planets and found that a lower expansion factor can possibly lead to a wider distribution of final planetary CMFs. This effect is more prominent when comparing results from low expansion factors (around 3 or 5) to high expansion factors (around 10 or 15) and likely comes from the smaller amount of disruptive collisions that occur in simulations with high expansion factors. Users must decide whether a higher expansion factor is worth the decrease in composition-changing collisions.

We then experimented with three, non-uniform distributions of CMFs in the initial disc of these simulations to see if they could be used to explain the high CMFs that have been observed in some exoplanets. In these distributions, an object's CMF depended on its placement in the initial disc, with the assumption that the inner edge of the disc is more iron-rich than the outer edge. We also set our model parameters to maximize the percentage of mantle material that can be in the ejecta from a collision as an additional aid to produce high CMF planets ($f\textsubscript{min}$ = 0.0 and $f\textsubscript{max}$ = 0.0). We found that fragments produced the widest range of CMFs, while planetary CMFs typically remained within the upper and lower bounds of the initial distribution. We also found that a larger planet mass leads to a planetary CMF that's closer to the final average CMF of the system. This trend could indicate that more massive planets must accrete more material from different parts of the disc to reach these sizes, which tends to average out their CMF towards the system mean. These results also showed a preference for planetary CMFs above the initial system average, which we determined was due to a large amount ejections of mantle-rich objects early on in these simulations.

With the non-uniform distributions and parameter settings described above, we obtained one planet that had a CMF greater than 0.9; this is the same planet that obtained a CMF of 0.037 with the DBCT parameters for the uniform disc. In addition, one distribution produced multiple planets with CMFs above 0.6. These planets likely began as embryos with CMFs above 0.6 and didn't experience many collisions during the course of their formation. These results imply two possible pathways for planets to obtain high CMFs: disruptive collisions near the end of a planet's formation that remove a large fraction of its mantle or forming from an already iron-rich planetary embryo.

We then studied lines in the top left panel of Fig. \ref{fig:three_final_cmfs} and plotted the final CMFs, masses, and semi-major axes of objects from all 50 simulations. After investigating, we found that the lines are likely artifacts that come from the use of a step-function distribution of core material in the initial disc with the DBCT values ($f\textsubscript{min}$ = 0.0 and $f\textsubscript{max}$ = 0.0) used when doing the analysis and does not have any physical significance. In the semi-major axis plot, we found the mass of a planet grows as its semi-major moves towards the central values for the initial disc, indicating that embryos near the middle region of the disc were able to accrete more material. We also found a lack of planets at semi-major axes past 2.0 AU, likely due to the large number ejections that occurred to embryos near the outer edge of the disc and the decreasing surface density of the initial disc at higher semi-major axes. Finally, in general, the trends between CMF and semi-major axis between the final planets and the initial objects in the disc were similar, showing that any radial mixing or erosive collisions that occurred during planet formation did not drastically alter the spread of core material in the system. 

Lastly, we analyzed the previous results produced by the DBCT by entering the same collisional data and non-uniform CMF distributions into the composition tracking code created by \citet{Childs2022} and comparing the outcomes of these two codes. Due to the inclusion of differentiation, the DBCT was able to produce many more fragments that lie outside the initial CMF range than the composition tracking code. However, the distributions of planetary CMFs produced by the two prescriptions was similar, though this result could change if an \Nbody\ simulation contains a high number of erosive collisions. Additionally, the planet with a CMF greater than 0.9 produced by the DBCT was not seen in the results for the composition tracking code and shows why differentiation is important to consider when looking at the compositional outcomes of giant impacts. 

Our DBCT is publicly available for use in future giant impact studies using \reb.   

\section*{Acknowledgements}

This work is supported in part by NSF grant AST-1910955.

\section*{Declaration of Competing Interest}

The authors declare that they have no known financial incentives or personal relationships that could have influenced the results presented in this manuscript.

\section*{Data Availability}

The Differentiated Body Composition Tracker with documentation is available at \href{https://github.com/Nofe4108}{https://github.com/Nofe4108}. The fragmentation module for \reb\ can be found at \href{http://github.com/ANNACRNN/REBOUND_fragmentation}{http://github.com/ANNACRNN/REBOUND\_fragmentation}. \reb\ is available at \href{https://rebound.readthedocs.io/en/latest/}{https://rebound.readthedocs.io/en/latest/}. The data used in this manuscript will be shared upon reasonable request to the corresponding author.


\bibliographystyle{elsarticle-harv} 
\bibliography{references}

\end{document}